\documentstyle[aps,eqsecnum,preprint,tighten,epsfig,amssymb]{revtex}

\begin{document}

\preprint{
\vbox{
\hbox{SNUTP-98-022}
\hbox{KOBE-FHD-98-1}
\hbox{nucl-th/9804043}
}}

\draft

%%%%%%%%%%%%%%%%%%%%% Title %%%%%%%%%%%%%%%%%%%%%%

\title{Photoproduction of $\bbox{\phi}$ mesons from the proton: \\
       Polarization observables and strangeness in the nucleon}

%%%%%%%%%%%%%%%%%%%% Authors %%%%%%%%%%%%%%%%%%%%%

\author{
Alexander I. Titov\,$^a$%
\footnote{E-mail address : {\tt atitov@thsun1.jinr.ru}},
Yongseok Oh\,$^b$%
\footnote{E-mail address : {\tt yoh@phya.snu.ac.kr}},
Shin Nan Yang\,$^c$%
\footnote{E-mail address : {\tt snyang@phys.ntu.edu.tw}},
and
Tosiyuki Morii\,$^d$%
\footnote{E-mail address : {\tt morii@kobe-u.ac.jp}}}

%%%%%%%%%%%%%%%%%%%% Addresses %%%%%%%%%%%%%%%%%%%%%

\address{$^a$\,Bogoliubov Laboratory of Theoretical Physics, JINR,
141980 Dubna, Russia\\
$^b$\,Research Institute for Basic Sciences and Department of Physics,
Seoul National University, Seoul 151-742, Korea\\
$^c$\,Department of Physics, National Taiwan University,
Taipei, Taiwan 10617, Republic of China\\
$^d$\,Faculty of Human Development,
Kobe University, 3-11 Tsurukabuto, Nada, Kobe 675, Japan}

\maketitle

%%%%%%%%%%%%%%%%%%%% Abstract %%%%%%%%%%%%%%%%%%%%%

\begin{abstract}
The polarization observables in $\phi$ meson photoproduction
is studied to probe the strangeness content of the nucleon.
In addition to the dominant diffractive production and the
one-pion-exchange process, we take into account the direct
knockout mechanism that arises from the possible hidden strangeness
content of the nucleon.
We find that some double polarization observables are very sensitive
to the strangeness content of the proton because of the different spin
structures of the amplitudes associated with different mechanisms.
This suggests that such measurements could be very useful in probing
the strangeness content in the proton.
The orbitally excited quark-cluster configurations in the proton are
included in the calculation and found to have little effect.

\end{abstract}

\pacs{PACS number(s): 13.88.+e, 24.70.+s, 25.20.Lj, 13.60.Le}

%%%%%%%%%%%%%%%%%%%% Introduction %%%%%%%%%%%%%%%%%%%%%%%%%

\section{Introduction}

The possible existence of hidden strangeness in the nucleon has recently
become one of the most controversial problems in nuclear/hadron physics.
Some analyses of the pion-nucleon sigma term \cite{DN86,GLS91a}, polarized
deep-inelastic lepton-proton scattering \cite{EMC89,SMC94,E143-95a},
and low energy elastic neutrino-proton scattering \cite{AACG87,BM91}
indicate a significant role of strange sea quarks in the nucleon
structure \cite{EGK89}.
However, it has also been argued that such experimental results could
be understood with little or null strangeness in the nucleon
\cite{AS89,Lip91}.

It will be interesting, therefore, to study other processes that might
be related directly to the strangeness content of the nucleon
\cite{KM88,McK89,HKPW91,Kap92,MDDP94,MTY96}.
One of them is $\phi$ meson production from the proton.
Since the $\phi$ meson is a nearly pure $s \bar s$ state because of
ideal mixing with the $\omega$ meson, its coupling to the proton is
suppressed through the OZI rule.
Then the idea is that we could extract information about the hidden
strangeness of the nucleon by studying the strange sea quark
contribution through the OZI evasion processes.
One example is $\phi$ production in proton--anti-proton annihilation.
Recent experiments on vector meson production
through $\bar pp$ annihilation at rest \cite{ASTERIX91,CBC95,OBELIX96}
report a strong violation of the OZI rule.
It can be accounted for by the presence of an intrinsic $s\bar s$
component in the nucleon wave function \cite{EGK89,Ger96}, which
contributes to the process through the rearrangement and shake-out
diagrams \cite{DF89,EKKS95,EK96,GFYY97}.
On the other hand, it was also claimed that this OZI violation could be
explained through modified meson exchange models \cite{LZL93,BL94}
without any strangeness content of the nucleon.

Another possibility is $\phi$ photo- and electro-production from proton
targets \cite{HKPW91}.
In this process, in addition to the vector-meson dominance model
(VDM), the contribution from the hidden strangeness of the proton arises
through the direct knockout process.
In Refs. \cite{HKW92,HFPY92}, Henley {\it et al.\/} calculated the
contribution from knockout process to $\phi$ electroproduction cross
section and found it comparable to that of VDM with an assumption
of a 10--20\% strange sea quark admixture in the proton wave function.
To arrive at this conclusion, they used nonrelativistic quark model wave
functions for the hadrons.
However, since the kinematical region of $\phi$ meson production is beyond
the applicability of the nonrelativistic quark model, the relativistic
corrections are expected to be important.
In Refs. \cite{TOY94,TYO97} we improved the calculations of Refs.
\cite{HKW92,HFPY92} with the use of a relativistic harmonic oscillator
quark model (RHOQM).
We found that the cross section of the direct knockout mechanism for
the electroproduction is comparable to that of VDM at moderately large
electron four-momentum transfer with less than 5\% admixture of strange
sea quarks in the proton.
However, it is not easy to disentangle the two mechanisms from the cross
section measurement because their respective contributions have similar
dependence on momentum transfer \cite{TYO97}.

To distinguish between the knockout and VDM processes, it was suggested
the difference in the spin structures of various amplitudes be exploited
\cite{HKW92,Lage93,TYO95,TOY97,Will98}.
In Ref. \cite{TOY97}, we showed that some double polarization observables
are indeed very sensitive to the hidden strangeness content of the proton.
We found that, with the use of RHOQM, the direct knockout process gives a
very distinct contribution to some of the double polarization observables
in $\phi$ photoproducton as compared to those of diffractive production
and one-pion-exchange (OPE) process.
A similar conclusion was drawn from the $\bar p p \to \bar \Lambda \Lambda$
process to distinguish between contributions from the hidden
strangeness of the nucleon and the effects from meson exchange
processes \cite{AEK95}. (See also Ref. \cite{PR98}.)
The one-pion-exchange process arises from the $\phi$-$\pi$-$\rho(\gamma)$
coupling.
Similar $\omega$-$\pi$-$\rho(\gamma)$ coupling gives non-negligible
effects in the $\omega$-meson production case \cite{JLMS77}.

In this paper, we extend our previous work to discuss other spin
observables in $\phi$ photoproduction and give the details which
were left out in Ref. \cite{TOY97}.
We also improve the VDM amplitude to take into account the gauge
invariance requirement within a quark-Pomeron interaction picture.
We further include, besides the lowest one, other configurations in
the 5-quark cluster model of the nucleon, which may give non-negligible
contribution to the nucleon spin \cite{Ger96}.

In Sec. II, we define the kinematical variables and briefly review the
definitions of general spin observables in terms of helicity amplitudes.
Section III is devoted to our model for $\phi$ photoproduction.
We include the diffractive and OPE production processes as well as the
direct knockout processes that arise from the hidden strangeness of the
nucleon.
The gauge invariance of VDM amplitude is discussed as well.
Our results for the spin observables are presented in Sec. IV along
with their dependence on the hidden strangeness content of the proton.
In Sec. V we discuss the role of orbitally excited quark-cluster
configurations in the nucleon wave function in $\phi$ photoproduction.
We find that their effect is not important.
Section VI contains summary and conclusion.
Some detailed discussions and expressions for the physical parameters are
given in Appendixes.

%%%%%%%%%%%%%%%%%%%  General Formalism  %%%%%%%%%%%%%%%%%%%

\section{Spin Observables and the Helicity Amplitudes}

\begin{figure}
\centering
\epsfig{file=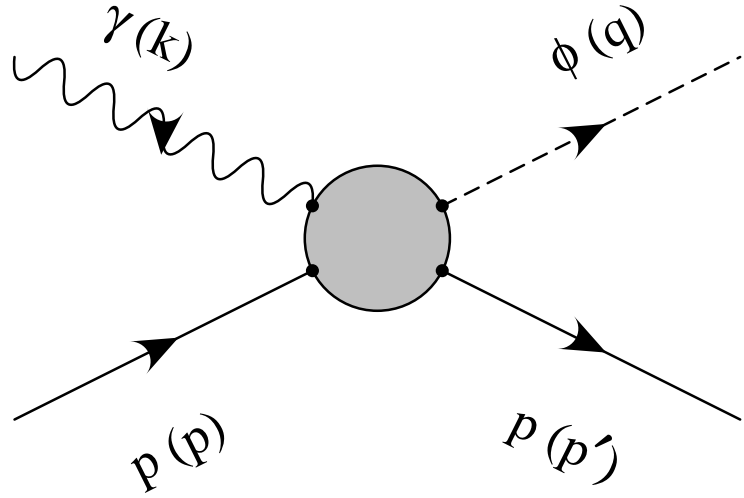, width=10cm}
\caption{Kinematics for $\phi$ meson photoproduction from the proton,
         $\gamma\,p \to \phi\,p$.}
\label{fig:phipr}
\end{figure}

\begin{figure}
\centering
\epsfig{file=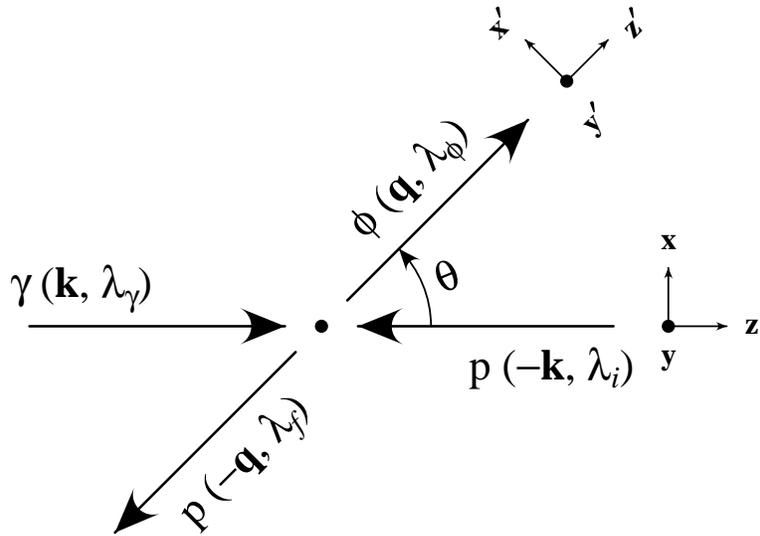, width=10cm}
\vskip 0.5cm
\caption{The coordinate system and kinematical variables for $\phi$ meson
photoproduction in the c.m. frame. $\theta$ is the scattering angle.}
\label{fig:CM}
\end{figure}

We first define the kinematical variables for $\phi$ photoproduction
from the proton, $\gamma + p \to \phi + p$, as shown in Fig. \ref{fig:phipr}.
The four-momenta of the incoming photon, outgoing $\phi$, initial
(target) proton, and final (recoil) proton are $k$, $q$, $p$, and $p'$,
respectively.
In the laboratory frame, we write
$k = (E_\gamma^L, {\bf k}_L)$, $q = (E_\phi^L, {\bf q}_L)$,
$p = (E_p^L, {\bf p}_L)$, and $p' = (E_{p'}^L, {\bf p}_L')$.
The variables in the c.m. system are written as
$k = (\nu, {\bf k})$, $q = (E_\phi, {\bf q})$,
$p = (E_p, -{\bf k})$, and $p' = (E_{p'}, -{\bf q})$, respectively, as in
Fig. \ref{fig:CM}.
We also define $t = (p - p')^2$ and $W^2 = (p+k)^2$ with $M_N$ the
nucleon mass, $M_\pi$ the pion mass, and $M_\phi$ the $\phi$ mass.
The differential cross section is given by
\begin{equation}
\frac{d\sigma}{d\Omega} = \rho_0 |T_{fi}|^2,
\end{equation}
where $\rho_0 =(M_N^2 |{\bf q}|) / (16 \pi^2 W^2 |{\bf k}|)$.

The general formalism for the spin observables of $\gamma + p \to \phi + p$
has been discussed extensively in the literature.
For completeness, we briefly review here the density matrix
formalism and refer the interested readers to Refs.
\cite{Ohl72,BLS80,Con94,PST96,STY96} for details.

To study the spin observables, it is useful to work with the helicity
amplitudes in the c.m. frame.
For polarized $\phi$ meson photoproduction, $\vec{\gamma} + \vec{p} \to
\vec{\phi} + \vec{p}$, the helicity amplitude takes the form,
\begin{eqnarray}
H_{\lambda_\phi,\lambda_f;\lambda_\gamma,\lambda_i} \equiv
\langle {\bf q}; \lambda_\phi, \lambda_f \,|\, T \,|\,
{\bf k}; \lambda_\gamma, \lambda_i \rangle,
\end{eqnarray}
where the variables and the coordinate systems are shown in
Fig. \ref{fig:CM} with $\lambda_\gamma$ $(=\pm 1)$,
$\lambda_\phi$ $(=0,\pm1)$, and $\lambda_{i,f}$ ($=\pm 1/2$) denoting
the helicities of the photon, $\phi$ meson, target proton, and recoil
proton, respectively.
We follow the Jacob-Wick phase convention \cite{BLS80,JW59} throughout
this paper.
In principle, there are $2 \times 2 \times 3 \times 2 = 24$ complex
amplitudes.
However, by virtue of parity invariance relation,
\begin{equation}
\langle {\bf q}; \lambda_\phi , \lambda_f | \, T \, | {\bf k};
\lambda_\gamma , \lambda_i \rangle =
(-1)^{\Lambda_f - \Lambda_i}
\langle {\bf q}; -\lambda_\phi , -\lambda_f \,|\, T \,|\,
{\bf k}; -\lambda_\gamma , -\lambda_i \rangle,
\label{phase}
\end{equation}
with $\Lambda_f = \lambda_\phi - \lambda_f$ and $\Lambda_i = \lambda_\gamma
- \lambda_i$, only 12 complex helicity amplitudes are independent.
We label them as \cite{PST96}
\begin{eqnarray}
H_{1,\lambda_\phi} &\equiv&
\langle \lambda_\phi, \, \lambda_f = + {\textstyle\frac12} \,|\, T\, |\,
\lambda_\gamma = 1, \, \lambda_i = - {\textstyle\frac12} \rangle,
\nonumber \\
H_{2,\lambda_\phi} &\equiv&
\langle \lambda_\phi, \, \lambda_f = + {\textstyle\frac12} \,|\, T\, |\,
\lambda_\gamma = 1, \, \lambda_i = + {\textstyle\frac12} \rangle,
\nonumber \\
H_{3,\lambda_\phi} &\equiv&
\langle \lambda_\phi, \, \lambda_f = - {\textstyle\frac12} \,|\, T\, |\,
\lambda_\gamma = 1, \, \lambda_i = - {\textstyle\frac12} \rangle,
\nonumber \\
H_{4,\lambda_\phi} &\equiv&
\langle \lambda_\phi, \, \lambda_f = - {\textstyle\frac12} \,|\, T\, |\,
\lambda_\gamma = 1, \, \lambda_i = + {\textstyle\frac12} \rangle.
\end{eqnarray}
The $\phi$-meson photoproduction amplitude can then be represented by a
$6 \times 4$ matrix ${\cal F}$ in helicity space;
\begin{equation}
{\cal F} \equiv \left( \begin{array}{rrrr}
H_{2,1} & H_{1,1} & H_{3,-1} & -H_{4,-1} \\
H_{4,1} & H_{3,1} & -H_{1,-1} & H_{2,-1} \\
H_{2,0} & H_{1,0} & -H_{3,0} & H_{4,0} \\
H_{4,0} & H_{3,0} & H_{1,0} & -H_{2,0} \\
H_{2,-1} & H_{1,-1} & H_{3,1} & -H_{4,1} \\
H_{4,-1} & H_{3,-1} & -H_{1,1} & H_{2,1} \end{array} \right).
\end{equation}

In actual calculations, sometimes it is easier to evaluate the matrix
elements in the nucleon spin space.
They are related to the helicity amplitude discussed above, in the
reference frame of Fig. \ref{fig:CM}, by
\begin{eqnarray}
H_{\lambda_\phi,\lambda_f;\lambda_\gamma,\lambda_i}
= (-1)^{1-\lambda_i-\lambda_f} \sum_{m_i,m_f}
d^{(1/2)}_{m_i,-\lambda_i}(0) d^{(1/2)}_{m_f,-\lambda_f}(\theta)
\langle \lambda_\phi, m_f \,|\, T \,|\, \lambda_\gamma, m_i \rangle.
\label{hel}
\end{eqnarray}
This expression reduces to that of Ref. \cite{FTS92} for the pseudoscalar
meson photoproduction process.

The differential cross section is given by the classical ensemble
average as
\begin{equation}
\frac{d\sigma}{d\Omega} = \rho_0 \, \mbox{Tr}\, (\rho_F).
\label{dsdom}
\end{equation}
The final state density matrix is
\begin{equation}
\rho_F = {\cal F} \rho_I {\cal F}^\dagger,
\end{equation}
where $\rho_I$ is the initial state density matrix,
\begin{equation}
\rho_I = \rho_\gamma \rho_N.
\end{equation}
The photon and proton density matrices, $\rho_\gamma$ and $\rho_N$,
are defined in Appendix A.
For example, in the unpolarized case where $\rho_\gamma=\rho_N=\frac12$,
we get
\begin{equation}
\frac{d\sigma}{d\Omega}^{(U)}
= \frac{\rho_0}{4} \, \mbox{Tr} \, ( {\cal F} {\cal F}^\dagger )
\equiv \rho_0 {\cal I} (\theta),
\end{equation}
which defines the cross section intensity ${\cal I}(\theta)$.

In general, any spin observable $\bar\Omega$ can be written as
\begin{equation}
\bar\Omega =
\frac{ \,\mbox{Tr}\,[ {\cal F} A_\gamma A_N {\cal F}^\dagger B_V B_{N'} ]}
{\,\mbox{Tr}\,({\cal F}{\cal F}^\dagger)},
\label{genspin}
\end{equation}
where $A_N$ denotes $(\bbox{1}_2, \bbox{\sigma}_N)$, which are elements
of the nucleon density matrix.
The explicit forms of $A_\gamma$, $B_{N'}$, and $B_V$ can be obtained
from the density matrices given in Appendix A.
Note that the dimensions of the matrices are ${\cal F} (6 \times 4)$,
$A_\gamma A_N (4 \times 4)$, ${\cal F}^\dagger (4 \times 6)$, and
$B_V B_{N'} (6 \times 6)$.

\subsection{Single polarization observables}

When only the incoming photon beam is polarized, we can define the
polarized beam asymmetry (analyzing power) $\Sigma_x$ as
\begin{equation}
\Sigma_x =
\frac{ \,\mbox{Tr}\,[ {\cal F} \sigma^x_\gamma {\cal F}^\dagger  ]}
{\,\mbox{Tr}\,({\cal F}{\cal F}^\dagger)}.
\end{equation}
If we define $\sigma^{(B,T;R,V)}$ for the cross section $d\sigma/d\Omega$
where the superscripts $(B,T;R,V)$ denote the polarizations of
(photon beam, target proton; recoil proton, produced vector-meson),
then the physical meaning of $\Sigma_x$ becomes clear through the relation,
\begin{equation}
\Sigma_x = \frac{ \sigma^{(\parallel,U;U,U)} - \sigma^{(\perp,U;U,U)}}
{ \sigma^{(\parallel,U;U,U)} + \sigma^{(\perp,U;U,U)}},
\end{equation}
where the superscript $U$ refers to an unpolarized particle and
$\parallel$ ($\perp$) corresponds to a photon linearly polarized
along the $\hat{{\bf x}}$ ($\hat{{\bf y}}$) axis.

Similarly, we can define the polarized target asymmetry $T$, recoil
polarization asymmetry $P$, and the vector-meson polarization asymmetry
$V$ as
\begin{eqnarray}
T_y &=& \frac{\,\mbox{Tr}\,( {\cal F} \sigma^y_N {\cal F}^\dagger )}
{\,\mbox{Tr}\, {\cal F} {\cal F}^\dagger)},
\nonumber \\
P_{y'} &=& \frac{ \,\mbox{Tr}\,( {\cal F}{\cal F}^\dagger \sigma^{y'}_{N'})}
{\,\mbox{Tr}\,( {\cal F}{\cal F}^\dagger)},
\nonumber \\
V_j &=& \frac{\,\mbox{Tr}\,({\cal F}{\cal F}^\dagger \Omega^V_j)}
{\,\mbox{Tr}\,({\cal F} {\cal F}^\dagger)},
\end{eqnarray}
where $\Omega^V_j$'s are given in Appendix A.
The explicit expressions for the single polarization observables can be
found in Appendix B.

\subsection{Double polarization observables}

There are six double polarization observables: Beam--Target (BT),
Beam--Recoil (BR), Target--Recoil (TR), Beam--Vector-meson (BV),
Target--Vector-meson (TV), and Recoil--Vector-meson (RV).
For example, we define the double polarization observables
$C_{ij}^{\rm BT}$ as%
\footnote{Our definitions of $C_{ij}$ are slightly different from those
of Ref. \cite{FTS92}. Our $C_{ij}$ corresponds to $C_{ji}$ of Ref.
\cite{FTS92}.}
\begin{equation}
C_{ij}^{\rm BT} = \frac{ \,\mbox{Tr}\,[ {\cal F} \sigma_\gamma^i
\sigma_N^j {\cal F}^\dagger]}{\,\mbox{Tr}\,( {\cal F}{\cal F}^\dagger)}.
\end{equation}
The physical meaning of $C_{zz}^{\rm BT}$ is then
\begin{eqnarray}
C_{zz}^{\rm BT} &=& \frac{ \,\mbox{Tr}\,[ {\cal F} \sigma_\gamma^z
\sigma_N^z {\cal F}^\dagger]}{\,\mbox{Tr}\,( {\cal F}{\cal F}^\dagger)}
\nonumber \\
&=& \frac{ \sigma^{(r,z;U,U)} - \sigma^{(r,-z;U,U)} }{ \sigma^{(r,z;U,U)} +
\sigma^{(r,-z;U,U)} },
\end{eqnarray}
where the superscript $r$ corresponds to a circularly polarized photon
beam with helicity $+1$, and $\pm z$ denotes the direction of the target
proton polarization.
Some of the double polarization observables are explicitly given in terms
of helicity amplitudes in Appendix B.
The complete list of double polarization observables can be found in,
e.g., Ref. \cite{PST96}.

Among the 290 possible (single, double, triple, and quadruple)
polarization observables, we will consider only a few of them including
longitudinal asymmetries.%
\footnote{We treat the cross section as a single polarization observable.
Although there are altogether 290 observables, only 24 of them are linearly
independent \cite{PST96}.}
For instance, we will not consider the $\phi$ meson tensor polarization
in the double polarization observables throughout this paper.

%%%%%%%%%%%%%%%%%%%%%%% Models %%%%%%%%%%%%%%%%%%%%%%%%%%%%

\section{The model for $\phi$ meson photoproduction}

To calculate the spin observables defined in the last Section, we need
to construct a model for the helicity amplitudes of $\phi$ photoproduction.
Our model includes the diffractive and OPE production processes and the
direct knockout of an $s \bar s$ (or $uud$) cluster in the proton.
We describe below the essential dynamics of each process and give the
resulting amplitude.

\subsection{Diffractive production}

\begin{figure}
\centering
\epsfig{file=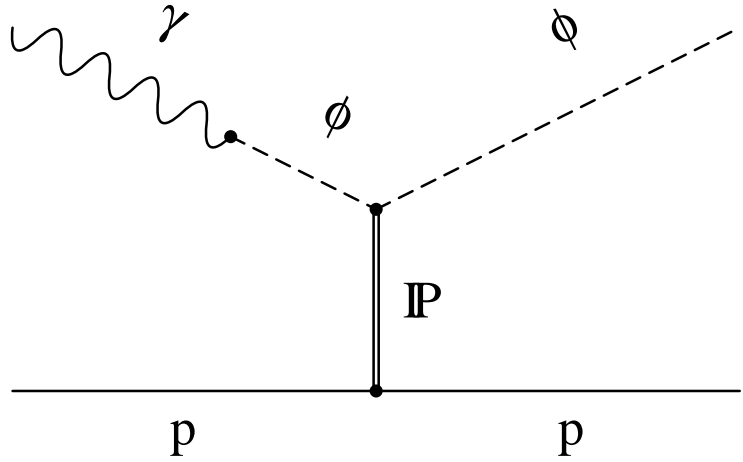, width=10cm}\\
\caption{Diffractive $\phi$ meson production within the
         vector-meson-dominance model through Pomeron exchange.}
\label{fig:VDM}
\end{figure}

In the VDM diffractive photoproduction \cite{CABD81,Leit78}, the
incoming photon first converts into vector mesons, i.e., the $\phi$-meson
in our case, and then this vector meson scatters diffractively from
the nucleon through Pomeron exchange, as shown in Fig. \ref{fig:VDM}.
Experimental observations for vector-meson production, small-$|t|$
elastic scattering, and diffractive dissociation indicate that Pomeron
behaves rather like a $C=+1$ isoscalar photon
\cite{DL84-86,PL96}.
A microscopic model for vector-meson photo- and electro-production
at high energy based on the Pomeron--photon analogy has been proposed
by Donnachie and Landshoff \cite{DL87a}, and the Pomeron could be
successfully described in terms of a non-perturbative two-gluon exchange
model \cite{Lage93,LN87,DL88-92,Cud90,Gol93,LM95}.

\begin{figure}
\centering
\epsfig{file=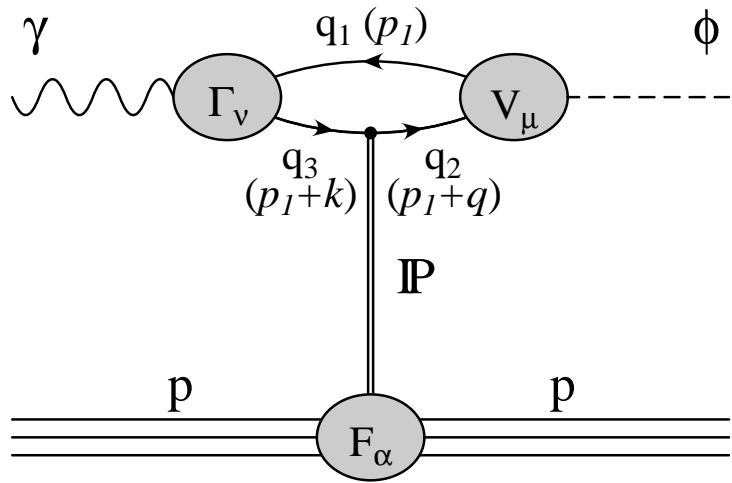, width=10cm}\\
\caption{Quark picture for the Pomeron exchange model of $\phi$
photoproduction.
The four-momenta of the quarks $q_{1,2,3}$ are given in parentheses.}
\label{fig:pomq}
\end{figure}

In our previous calculation \cite{TOY97}, we used the vector-meson
dominance model with Pomeron-photon analogy within the hadron-Pomeron
interaction picture, which is expected to be valid in the low energy
region.
In this paper, we employ a microscopic model for the VDM.
In this approach, the incoming photon first converts into a quark and
antiquark pair, which then exchanges a Pomeron%
\footnote{We do not consider two-gluon-exchange model for the Pomeron
in this work.}
with one of the quarks in the proton before recombining into an outgoing
$\phi$ meson, as depicted in Fig. \ref{fig:pomq}.
(See, e.g., Ref. \cite{PL96}.)
In terms of $\phi$ (photon) polarization vector $\varepsilon_{\phi}$
($\varepsilon_\gamma$), the invariant amplitude of the diffractive
production can be written as
\begin{equation}
T_{fi}^{\rm VDM} = i T_0  \varepsilon^*_{\phi\mu} {\cal M}^{\mu\nu}
\varepsilon_{\gamma\nu},
\label{T-VDM}
\end{equation}
with
\begin{equation}
{\cal M}^{\mu\nu} = {\cal F}_\alpha \Gamma^{\alpha,\mu\nu},
\end{equation}
where ${\cal F}_\alpha$ describes the Pomeron-nucleon vertex and
$\Gamma^{\alpha,\mu\nu}$ is associated with the Pomeron--vector-meson
coupling which is related to the $\gamma \to q \bar q$ vertex $\Gamma_\nu$
and the $q\bar q \to \phi$ vertex $V_\mu$, as shown in Fig. \ref{fig:pomq}.
The dynamics of the Pomeron-hadron interactions is contained in $T_0$.

To determine the explicit forms of the vertices, we have to rely on some
model assumptions.
Based on the Pomeron-photon analogy, the quark-quark-Pomeron vertex, i.e.,
$q_2 q_3 \mathbb{P}$ vertex in Fig. \ref{fig:pomq}, is assumed to
be $\gamma_\mu$.
Accordingly, we also have
\begin{equation}
{\cal F}_\alpha = \bar u (p') \gamma_\alpha u(p),
\label{Fal}
\end{equation}
where $u(p)$ is the Dirac spinor of the proton with momentum $p$.
The factor $N_q$ of the number of quarks in the proton can be
absorbed into $T_0$.

With the assumptions used in Ref. \cite{PL96}, namely, (i) quarks
$q_1$ and $q_2$ which recombine into a $\phi$-meson are almost on-shell
and share equally the 4-momentum of the outgoing $\phi$, i.e., the
nonrelativistic wave function assumption, (ii) quark $q_3$, which is
between photon and Pomeron, is far off-shell, and
(iii) $\Gamma_\nu \propto \gamma_\nu$ and $V_\mu \propto \gamma_\mu$,
the loop integral in Fig. \ref{fig:pomq} can be easily carried out
to give
\begin{equation}
\bar\Gamma^{\alpha,\mu\nu} \propto 2 \, \mbox{Tr}\,
\left\{ \gamma^\mu (\not\hskip-0.7mm\!{p}_1 + M_s ) \gamma^\nu
(\not\hskip-0.7mm\!{p}_1 + \not\hskip-0.7mm\!{k} + M_s ) \gamma^\alpha
(\not\hskip-0.7mm\!{p}_1 + \not\hskip-0.7mm\!{q} + M_s ) \right\},
\end{equation}
where $M_s$ is the $s$ quark mass and $p_1$ is the 4-momentum of the
quark $q_1$.
Explicit calculation leads to
\begin{equation}
\bar\Gamma^{\alpha,\mu\nu} =
2k^\alpha g^{\mu\nu}- \frac{2}{q^2} k^\alpha q^\mu q^{\nu}
- 2 g^{\alpha\nu} (k^\mu - q^\mu\,\frac{k \cdot q}{q^2})
+ 2 (k^\nu - q^\nu) (g^{\mu\alpha}-\frac{q^\alpha q^\mu}{q^2}).
\label{Gam-0}
\end{equation}
Inspection of Eq. (\ref{Gam-0}), however, shows that the last term breaks
the gauge invariance%
\footnote{This problem has also been discussed in Refs. \cite{PL96,CRHLD}.
To cure this problem, it was suggested that the quark-gluon structure of
the Pomeron in QCD be described in a consistent way \cite{CRHLD}, or
the correct off-shell structure of the electromagnetic interaction of the
dressed quarks be taken into account in constituent quark models \cite{PL96}.
However, further detailed discussion on this topic is beyond the scope of
this work.}
so that ${\cal M}^{\mu\nu} k_\nu \neq 0$.
This arises from the simple assumption about $\Gamma_\nu$ and a more
realistic modification of $\Gamma_\nu$ is needed to fix this problem
\cite{PL96}.
To have a gauge invariant amplitude, here we simply remove
the gauge non-invariant terms by multiplying the projection operator
${\cal P}_{\mu\nu}$ from both the left- and right-hand sides of
$\bar\Gamma^{\alpha,\mu\nu}$ \cite{FS69}, i.e.,
\begin{equation}
\bar\Gamma^{\alpha,\mu\nu} \to \Gamma^{\alpha,\mu\nu} =
{\cal P}^{\mu\mu'} \bar\Gamma^{\alpha}_{\mu'\nu'} {\cal P}^{\nu'\nu},
\end{equation}
where
\begin{equation}
{\cal P}_{\mu\nu} = g_{\mu\nu} - \frac{1}{k \cdot q} k_\mu q_\nu.
\end{equation}
It leads to a modified $\Gamma^{\alpha,\mu\nu}$ as
\begin{eqnarray}
\Gamma^{\alpha,\mu\nu} &=& (k+q)^\alpha g^{\mu\nu} - 2 k^\mu g^{\alpha\nu}
\nonumber \\ && \mbox{}
+ 2 \left[ k^\nu g^{\alpha\mu}
  + \frac{q^\mu}{q^2} ( k \cdot q g^{\alpha\nu} - k^\alpha q^\nu
                       - q^\alpha k^\nu)
  - \frac{k^2 q^\nu}{q^2 k \cdot q}
                  ( q^2 g^{\alpha\mu} - q^\alpha q^\mu )
\right]
\nonumber \\ && \mbox{}
+ (k-q)^\alpha g^{\mu\nu}.
\label{Gamalmunu}
\end{eqnarray}
Note that although the third term within the square brackets in Eq.
(\ref{Gamalmunu}) is essential to ensure the gauge invariance
it does not play any role in $\phi$ photoproduction because
$ q \cdot \varepsilon_\phi = k \cdot \varepsilon_\gamma = 0$ and
$k^2 = 0$ in photoproduction.
The last term also does not contribute because
${\cal F} \cdot k = {\cal F} \cdot q$.
Equation (\ref{Gamalmunu}) completes our prescription for the spin structure
of VDM amplitude.
This should be compared with the $\tilde\Gamma^{\alpha,\mu\nu}$ that was
used in Ref. \cite{TOY97},
\begin{equation}
\tilde\Gamma^{\alpha,\mu\nu} =
( k + q )^\alpha g^{\mu\nu} - k^\mu g^{\alpha\nu} - q^\nu g^{\alpha\mu},
\label{VDMold}
\end{equation}
which was obtained by gauging the massive vector-field Lagrangian in the
usual way \cite{BD,LY62} for the $\phi \phi \mathbb{P}$ vertex.
Note that $\tilde\Gamma^{\alpha,\mu\nu}$ is obtained within the
hadron-Pomeron interaction picture while we attempt to use a microscopic
quark-Pomeron interaction scheme instead in this paper.
Note the similarity between Eqs. (\ref{Gamalmunu}) and (\ref{VDMold})
as well.
More detailed discussion on the comparison of $\Gamma^{\alpha,\mu\nu}$
with $\tilde\Gamma^{\alpha,\mu\nu}$ is given in Appendix C together
with the gauge invariance of $\tilde\Gamma^{\alpha,\mu\nu}$.

The factor $T_0$ in Eq. (\ref{T-VDM}) includes the dynamics of the
Pomeron-hadron interaction.
We use the form and parameters of $T_0$ determined in Ref. \cite{BHKK74},
which reads
\begin{equation}
\left( \frac{d\sigma}{dt} \right)_{\rm VDM} = \sigma_\gamma (W) b_\phi
\exp( - b_\phi | t - t_{\rm max} | ),
\label{dcsVDM}
\end{equation}
with $b_\phi = 4.01$ GeV$^{-2}$ and $\sigma_\gamma (W) = 0.2$ $\mu$b around
$W = 2 \sim 3$ GeV.%
\footnote{
There are two comments concerning the parameters. First, these parameters
may be dependent on the energy scale. However, for our present
qualitative study we will assume constant values for them
at $W = 2 \sim 3$ GeV throughout this paper.
Second, the parameters are determined by fitting the formula
(\ref{dcsVDM}) to the experimental data, so the contributions
from the knockout and OPE processes are neglected.
However, as we will see, these mechanisms of the $\phi$ photoproduction
are suppressed compared with that of the VDM and the use of this
parameter set is justified.}
This normalizes the amplitude $T_0$ and explicitly we have
\begin{equation}
T_0 = \frac{W^2 - M_N^2}{M_N {\cal N}}
\sqrt{ 4\pi \sigma_\gamma(W) b_\phi }
\exp( - {\textstyle\frac12} b_\phi | t - t_{\rm max} | ),
\label{Tform}
\end{equation}
where
\begin{eqnarray}
t_{\rm max} = |t|_{\rm min} = 
2 M_N^2 - 2 E_p E_{p'} + 2 | {\bf k}| |{\bf q}|,
\end{eqnarray}
and the normalization constant ${\cal N}$ reads
\begin{eqnarray}
{\cal N}^2 &=& \frac{2}{M_N^2 M_\phi^2}
\{ k \cdot p \left[ k \cdot p \, M_\phi^2 + (k \cdot q)^2 \right]
 + 2 k \cdot p\,  k \cdot q [ p \cdot q -2 M_\phi^2 ]
\nonumber \\ && \mbox{} \qquad
- (k \cdot q)^2 [ p \cdot q+M_N^2 ] \}.
\end{eqnarray}

It is now straightforward to obtain the VDM helicity amplitude as
\begin{eqnarray}
H^{\rm VDM}_{\lambda_\phi,\lambda_f;\lambda_\gamma,\lambda_i} &=&
\sum_{m_i,m_f} d^{(1/2)}_{m_f,\lambda_f} (\pi+\theta)
d^{(1/2)}_{m_i,\lambda_i} (\pi)
T^{\rm VDM}_{\lambda_\phi,m_f;\lambda_\gamma,m_i}
\label{helamp-VDM}
\end{eqnarray}
where
\begin{eqnarray}
T^{\rm VDM}_{\lambda_\phi,m_f;\lambda_\gamma,m_i} &=& i
C T_{0} \Biggl\{ \Bigl[ (1+ \alpha\alpha'\cos\theta)\,
({\cal V}^{0}-{\cal W}^{0}) \nonumber\\
&& \qquad \qquad \mbox{}
-a^z({\cal V}^{z}-{\cal W}^{z})
+a^x{\cal W}^{x} - 2m_i\,b^x\, \mbox{Im} \,{\cal W}^{y}
\Bigr]\,
\delta_{m_i\,m_f}\nonumber\\
&& \mbox{} \qquad
+2m_i\,
\Bigl[\alpha\alpha' \sin\theta ({\cal V}^{0}-{\cal W}^{0})
\nonumber
\\
&& \qquad \qquad \mbox{}
-b^x ({\cal V}^{z} -{\cal W}^{z})-b^z{\cal W}^{x}
+\frac{1}{2m_i}\,b^z\,\mbox{Im}\,{\cal W}^{y}\,
\Bigr]\delta_{m_i\,-m_f} \Biggr\},
\label{hel-VDM}
\end{eqnarray}
with $C=\sqrt{(\gamma_p+1)(\gamma_p'+1)}/2$ and $\theta$ is the c.m.
scattering angle.
Definitions for the other variables and their detailed derivation are
given in Appendix D.
Close inspection of this amplitude shows that at small $|t|$
(or $\theta\to 0$), the dominant part, namely the $(k+q)^\alpha g^{\mu\nu}$
term in $\Gamma^{\alpha,\mu\nu}$, has the spin/helicity conserving
form as known in the conventional VDM amplitude,
\begin{eqnarray}
T^{\rm VDM}_{\lambda_\phi,m_f;\lambda_\gamma,m_i}\simeq
-2i|{\bf k}| C T_{0}(1+ \alpha\alpha')\,\delta_{\lambda_\phi\,\lambda_\gamma}
\equiv -i M^{\rm VDM}_0\,\delta_{\lambda_\phi\,\lambda_\gamma}\,
\delta_{m_i\,m_f},
\label{hel-VDM1}
\end{eqnarray}
while the spin-flip part is suppressed.
Note also that $T^{\rm VDM}$ is purely imaginary.

\subsection{One-pion-exchange in $\phi$ photoproduction}

\begin{figure}
\centering
\epsfig{file=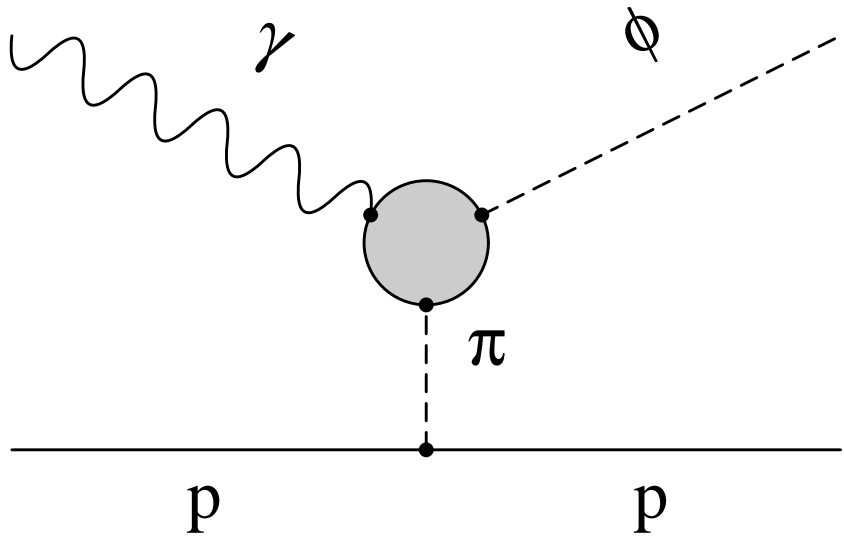, width=0.7\hsize}
\caption{One pion exchange process in $\phi$ photoproduction.}
\label{fig:OPE}
\end{figure}

At low photon energy, one-pion-exchange diagram (Fig. \ref{fig:OPE})
gives non-negligible contribution.
This may be regarded as a correction to the VDM process \cite{JLMS77}.

\begin{figure}
\centering
\epsfig{file=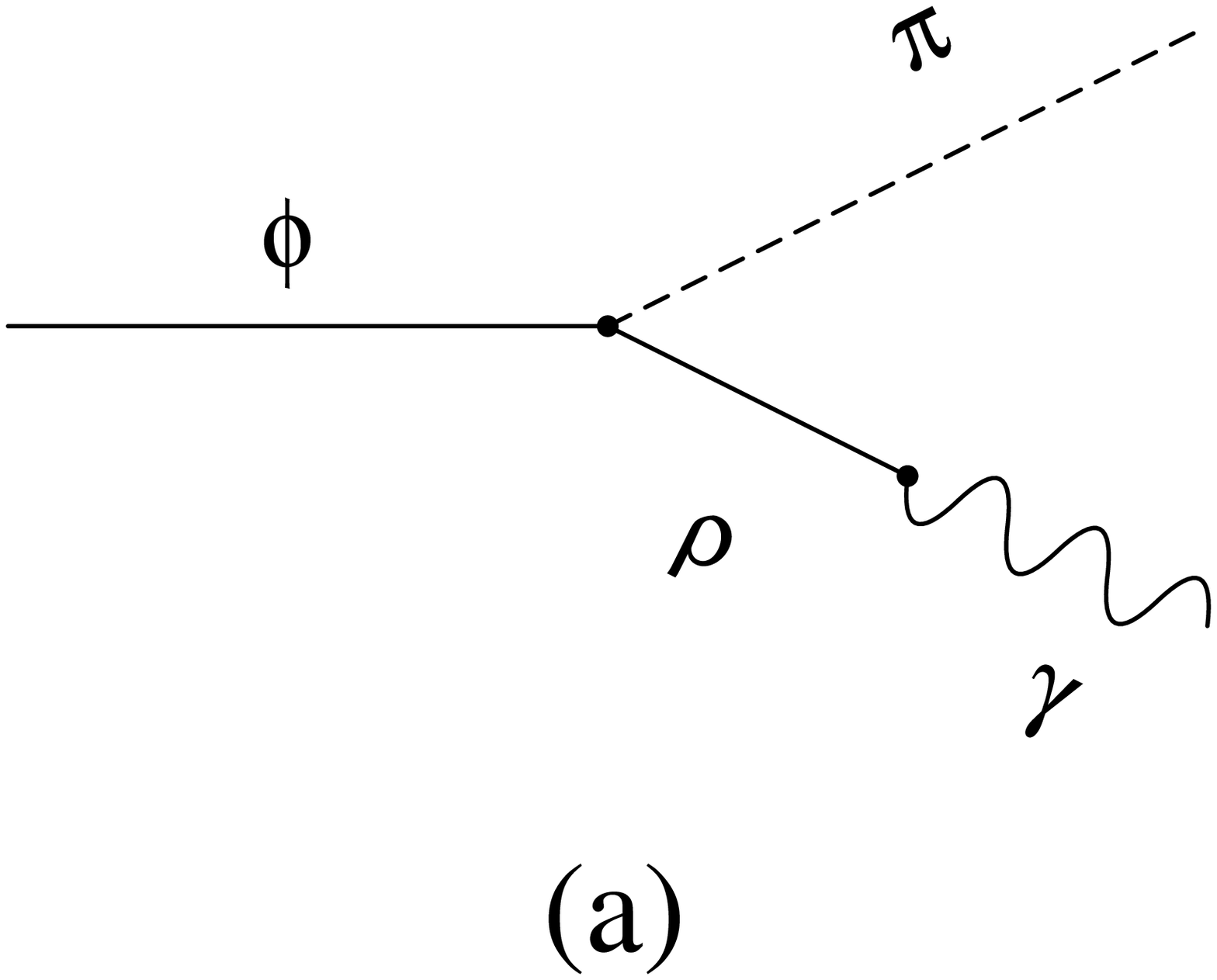, width=7cm} \qquad
\epsfig{file=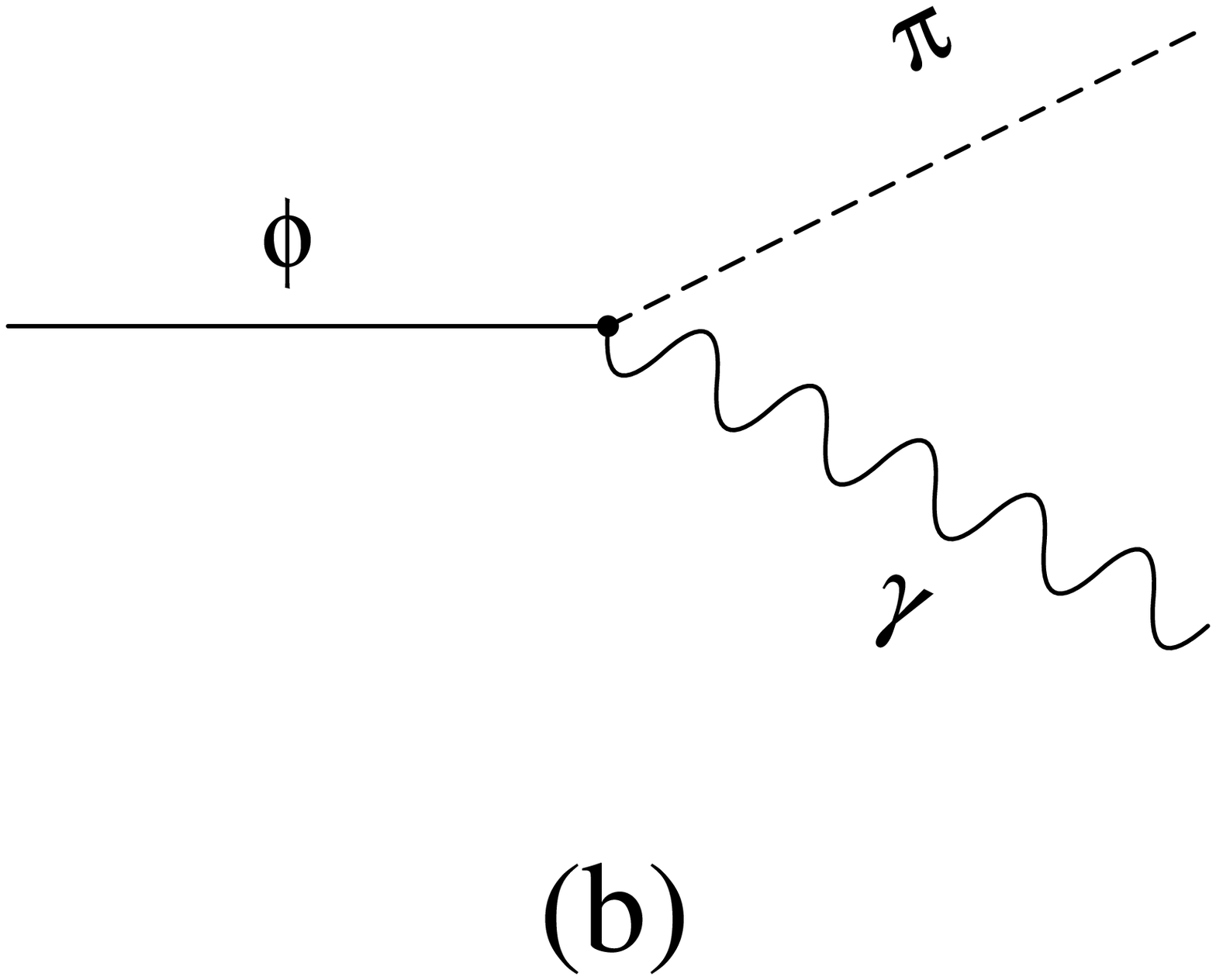, width=7cm}
\caption{Two possible mechanisms of $\phi\to\gamma\pi$ decay.}
\label{fig:phidec}
\end{figure}

The effective Lagrangian for the $\phi\gamma\pi$ interaction has
the form,
\begin{eqnarray}
{\cal L}_{\phi\gamma\pi} = \tilde g_{\phi\gamma\pi}
\epsilon^{\mu\nu\alpha\beta}
\partial_\mu \phi_\nu \partial_\alpha A_\beta \pi^0,
\label{L_phigammapi}
\end{eqnarray}
where $A_\beta$ is the photon field.
The effective coupling constant $\tilde g_{\phi\gamma\pi}$ can be
estimated through the decay width of $\phi \to \gamma\pi$, which reads
\begin{eqnarray}
\Gamma( \phi\to\gamma\pi) = \frac{1}{96\pi}
\frac{(M_\phi^2 - M_\pi^2)^3}{M_\phi^3} \tilde g_{\phi\gamma\pi}^2.
\end{eqnarray}
{}From the empirical value of $\Gamma( \phi\to\gamma\pi^0) =
5.8 \times 10^{-6} \mbox{ GeV}$, we get
$\tilde g_{\phi\gamma\pi} = 0.042 \mbox{ GeV}^{-1}$.
A remark is needed here concerning this estimate.
The blob in Fig. \ref{fig:OPE} contains two processes as shown in Fig.
\ref{fig:phidec}.
In addition to the VDM-like process of Fig. \ref{fig:phidec}(a),
there is another Gell-Mann--Sharp--Wagner type diagram shown in
Fig. \ref{fig:phidec}(b).
In the pure VDM, the decay process is completely dominated by
Fig. \ref{fig:phidec}(a) and there is no contact term.
However, this pure VDM diagram gives
\begin{eqnarray}
\Gamma( \phi\to\gamma\pi)_{\rm VDM} = \frac{\alpha_e}{24}
\frac{g_{\phi\rho\pi}^2 (M_\phi^2 - M_\pi^2)^3}
{M_\phi^3 f_\rho^2} = 1.65 \times 10^{-5} \mbox{ GeV},
\end{eqnarray}
with the $\rho$-meson decay constant $f_\rho$ ($=5.04$),
$\alpha_e = e^2/4\pi$, and $g_{\phi\rho\pi} = 1.19$ GeV$^{-1}$.
Thus the pure VDM overestimates the decay width by a factor
of 3 and we have to allow for the contact term of Fig. \ref{fig:phidec}(b)
to fit the experimental decay width.
However, since the two transition amplitudes of Fig. \ref{fig:phidec}
have the same structure, we combine the two processes into one term as in
Eq. (\ref{L_phigammapi}) with an effective coupling constant
$\tilde g_{\phi\gamma\pi}$.

For the $NN\pi$ interaction, one can use either pseudoscalar or
pseudovector coupling which are equivalent at the tree level.
For definiteness we use the pseudoscalar coupling of the form
\begin{equation}
{\cal L}_{PS} = -i g_{\pi NN} \bar N \gamma_5
                 \bbox{\tau} \cdot \bbox{\pi} N,
\end{equation}
with ${g_{\pi NN}^2}/{4 \pi} = 14.3$.

To include the off-shell effects, each vertex in Fig. \ref{fig:OPE} has
to be modified with a form factor.
We follow Ref. \cite{JLMS77} and use the Benecke-D{\"u}rr form factors
\cite{BD68} in which the $\pi N N$ form factor $F_N$ and the
$\phi\gamma\pi$ form factor $F_\phi$ are parameterized as
\begin{eqnarray}
F_N = \frac{1 + (2.9)^2 Q_N^2}{1 + (2.9)^2 Q_{NT}^2},  \qquad
F_\phi = \frac{U(2.3 Q_F)}{U(2.3 Q_T)} \left( \frac{Q_T}{Q_F} \right)^2,
\label{2_3}
\end{eqnarray}
where $Q_N$ ($Q_{NT}$) is the on-shell (off-shell) $\pi$-$N$ c.m.
momentum and $Q_T$ ($Q_F$) is the momentum of the on-shell
(off-shell) pion in the $\phi$ rest frame \cite{Wolf69}, respectively,
\begin{eqnarray}
&& Q_N^2 = \frac{M_\pi^2 ( M_\pi^2 - 4 M_N^2)}{4 M_N^2},  \qquad
Q_{NT}^2 = \frac{t(t-4 M_N^2)}{4M_N^2},
\nonumber \\ &&
Q_T = \frac{1}{2M_\phi} (M_\phi^2 - M_\pi^2), \qquad
Q_F = \frac{1}{2M_\phi} (M_\phi^2 - t).
\end{eqnarray}
And $U(x)$ is given as
\begin{eqnarray}
U(x) = \frac{1}{2x^2} \left[ \frac{2x^2+1}{4x^2} \log (4x^2 + 1) - 1 \right].
\end{eqnarray}
Before using these form factors, one should be careful with the
use of factor $2.3$ in $F_\phi$ of Eq. (\ref{2_3}) since this factor
is determined for the $\omega\gamma\pi$ coupling \cite{JLMS77}.
However, since we do not have enough data for the $\phi$ meson case, we
will use this value in our qualitative study on $\phi$ photoproduction.

The $T$ matrix element of the OPE process then reads
\begin{eqnarray}
T_{fi}^{\rm OPE} = \frac{i}{t - M_\pi^2} g_{NN\pi}
\tilde g_{\phi\gamma\pi} W^F_{m_f,m_i} W^B_{\lambda_\phi,\lambda_\gamma},
\label{TOPE}
\end{eqnarray}
where the coupling constants contain the Benecke-D{\"u}rr form factors
and
\begin{eqnarray}
W^F_{m_f,m_i} = \bar u(p') \gamma_5 u(p),  \qquad
W^B_{\lambda_\phi,\lambda_\gamma} = \epsilon^{\mu\nu\alpha\beta}
q_\mu k_\alpha \varepsilon_{\phi\nu} \varepsilon_{\gamma\beta}.
\end{eqnarray}
Direct calculation of $W^F$ and $W^B$ gives
\begin{eqnarray}
W^F_{m_f,m_i} &=& C\,
\left[ 2 m_f (\alpha' \cos\theta - \alpha) \delta_{m_f\,m_i}
      - \alpha' \sin\theta \delta_{m_f\,-m_i}
\right],
\nonumber \\
W^B_{\lambda_\phi,\lambda_\gamma} &=& i E_\gamma
\left[ \lambda_\gamma (E_\phi - |{\bf q}| \cos\theta)
        \bbox{\varepsilon}_\phi \cdot \bbox{\varepsilon}_\gamma
      + \frac{|{\bf q}| \sin\theta}{\sqrt{2} M_\phi}
        (|{\bf q}| - |{\bf q}| E_\phi \cos\theta) \delta_{\lambda_\phi\,0}
\right.\nonumber \\ && \qquad \left. \mbox{}
      - \frac{\lambda_\phi |{\bf q}| \sin^2\theta}{2}\right],
\end{eqnarray}
where
\begin{equation}
\bbox{\varepsilon}_\phi \cdot \bbox{\varepsilon}_\gamma
=[1+(\frac{E_\phi}{M_\phi}-1) \delta_{\lambda_\phi\,0}]
d^{(1)}_{\lambda_\gamma,\lambda\phi}(\theta),
\end{equation}
and $\alpha$ and $\alpha'$ are given in Appendix D.
Note also that the OPE amplitude is purely real.
This implies that it does not interfere with the knockout amplitudes in
the differential cross section as we shall see below.

One may also consider the $t$-channel $\eta$-meson exchange instead of
pion.
In fact, the decay width of $\phi \to \gamma\eta$ is about $5.58 \times
10^{-5}$ GeV, which is larger than $\Gamma(\phi\to \gamma\pi)$ by an
order of magnitude.
This gives us the large $\phi\gamma\eta$ coupling constant
$g_{\phi\gamma\eta} = 0.218$ GeV$^{-1}$ as compared to
$g_{\phi\gamma\pi} = 0.042$ GeV$^{-1}$.
However, we should also consider the $\eta NN$ coupling.
By assuming SU(3) flavor symmetry, one obtains $g_{\eta NN} / g_{\pi^0
NN} = \frac{1}{\sqrt3} \cdot (D-3F)/(D+F) \simeq -0.19 \sim -0.35$
using $F/D = 0.5 \sim 2/3$, and we find that the product of the coupling
constants in $\eta$-exchange diagram is of the same order of magnitude as
that of OPE.
Nevertheless, because of its heavier mass, the $\eta$-meson exchange amplitude
is expected to be smaller than that of OPE at least in the forward
scattering region.
There can also be cancellation between the two because
$g_{\eta NN} / g_{\pi^0 NN} < 0$.
In this work, therefore, we will not consider the $\eta$-meson exchange
diagram in $\phi$ photoproduction.%
\footnote{Furthermore, since this one boson exchange amplitude is purely real,
its contribution to the double polarization observables is expected to be
negligible. See Eq. (\ref{CBT00}).}

\subsection{Direct knockout production}

\begin{figure}
\centerline{
\epsfig{file=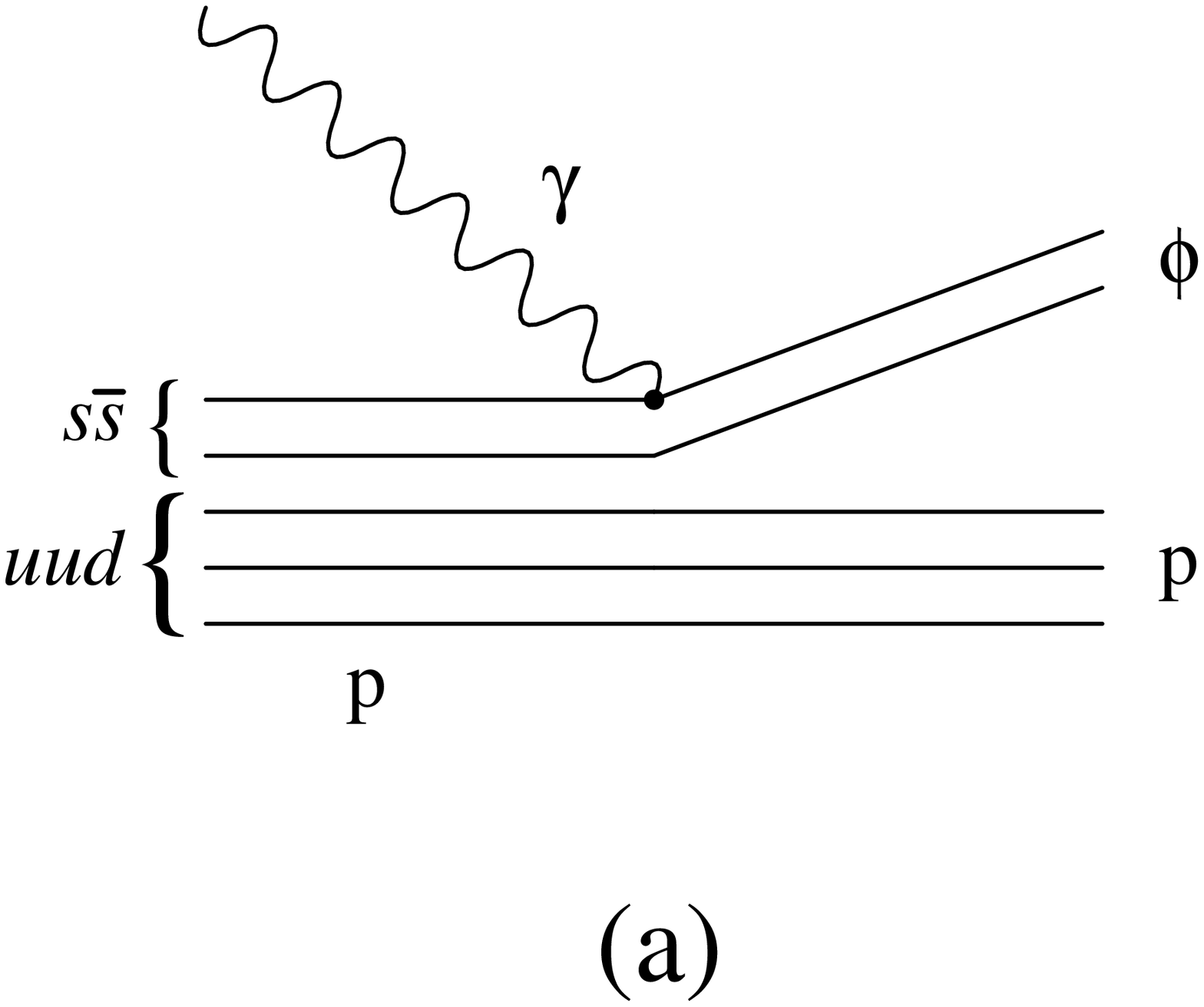, width=7cm} \qquad
\epsfig{file=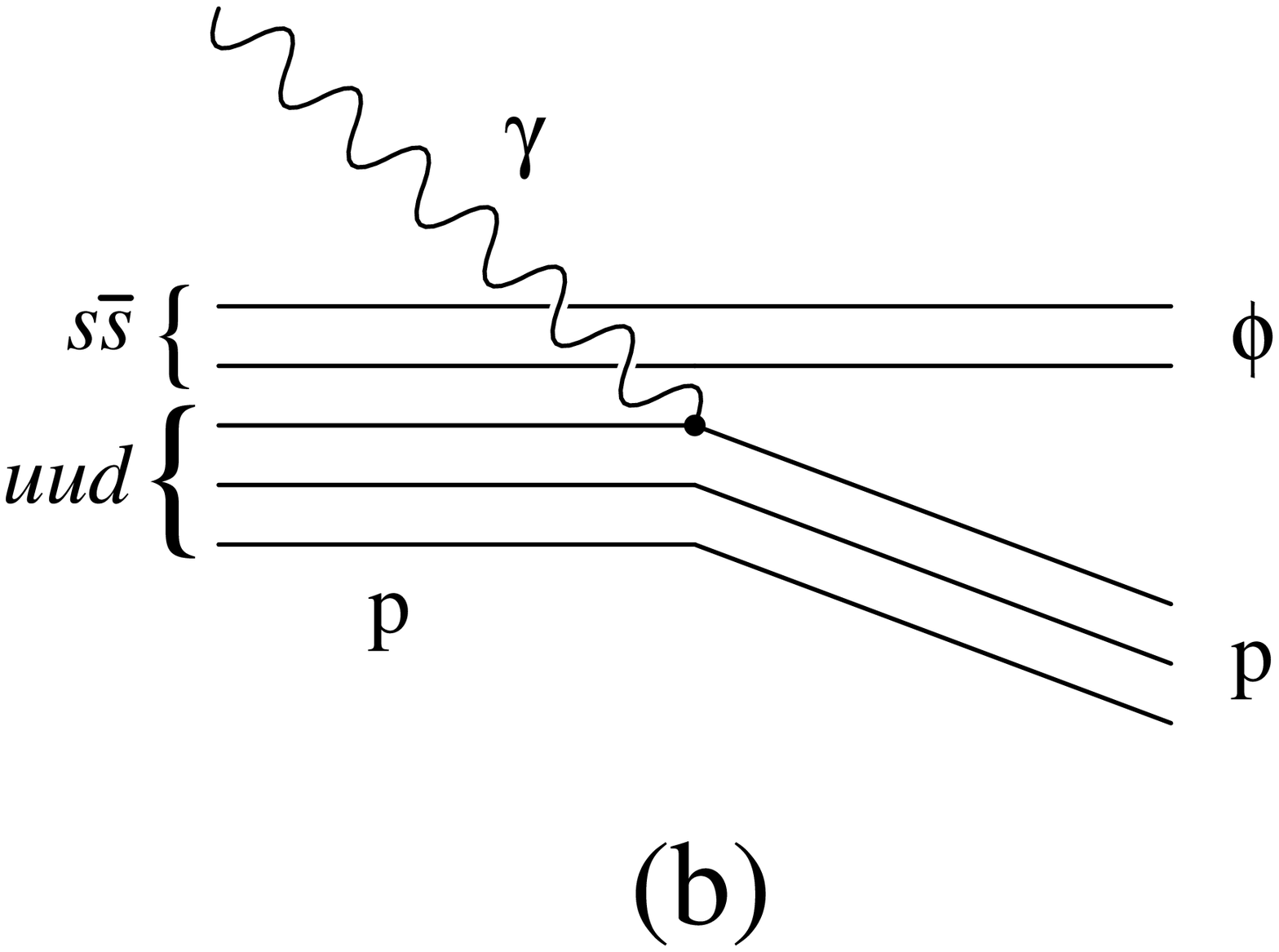, width=7cm}}
\caption{(a) $s\bar s$-knockout and (b) $uud$-knockout contributions to
         $\phi$ meson photoproduction.}
\label{fig:KO}
\end{figure}

When the incoming photon interacts with the 5-quark component of the
proton, we have an additional process called direct knockout as shown
in Fig. \ref{fig:KO}.
This process can be classified, according to the struck quark-cluster,
into $s\bar s$- and $uud$-knockout.
In order to investigate the effects from the hidden strangeness content
of the proton in $\phi$ photoproduction, we parameterize
the proton wave function in Fock space as
\begin{equation}
|\, p \rangle = A_0 |\, uud \rangle + \sum_X A_X |\, uud X \rangle
+ \sum_X B_X |\, uuds\bar s X \rangle,
\label{proton}
\end{equation}
where $X$ denotes any combination of gluons and light quark pairs of $u$
and $d$ quarks. Our aim is to estimate $|B_X|^2$ by isolating the OZI
evasion processes. Ellis {\it et al.\/} \cite{EKKS95} estimated it to be
1--19\% from an analysis of $p \bar p$ annihilation.
{}From the $\phi$ electroproduction process, Henley {\it et al.\/}
\cite{HKW92} claimed that its theoretical upper-bound would be 10--20\%.
We improved their prediction by employing a relativistic quark model
\cite{TOY94,TYO97}, and showed that the upper-bound could be lowered
to 3--5\%.

For simplicity and for our qualitative study, we approximate the proton
wave function (\ref{proton}) as
\begin{equation}
|\, p \rangle = A |\, uud \rangle + B |\, uud s\bar s \rangle.
\end{equation}
This parameterization of the nucleon wave function can be justified in
our case of $\phi$ production as argued in Refs.
\cite{HKW92,HFPY92,TOY94,TYO97}.
To compensate for the negative parity of the $s \bar s$ cluster,
only the odd orbital excitations in the wave function of relative motion
between $uud$ and $s\bar s$ clusters are allowed.
In principle, there are two more configurations when we consider the
first orbital excitation of the quark clusters:
either $s\bar s$-cluster or $uud$-cluster is orbitally excited.
In this Section, we consider only the $s\bar s$ clusters with
$j_{s\bar s}^P = 0^-$ and $1^-$, where $j_{s\bar s}^P$ stands for the
spin of an $s \bar s$ cluster of parity $P$, and leave the study of the
other quark-cluster configurations to Sec. V.
The proton wave function can then be expressed as
\begin{eqnarray}
|p\rangle = A | [uud]^{1/2} \rangle +
\sum_{j_{s\bar s} = 0,1; \, j_c} b_{j_{s\bar s}}
| \left[ \bbox{[} [uud]^{1/2} \otimes [{\bf L}]\bbox{]}^{j_c}
 \otimes [s\bar s]^{j_{s\bar s}} \right]^{1/2}
 \rangle ,
\label{protonwf}
\end{eqnarray}
where the superscripts $1/2$ and $j_{s\bar s}$ denote the spin of each
cluster and $(b_0, b_1)$ correspond to the amplitudes of the $s\bar s$
cluster with spin $0$ and $1$, respectively.
The strangeness admixture of the proton, $B^2$, is then defined to be
$ \sum |b_{j_{s\bar s}}|^2$, which is constrained to $A^2 + B^2 = 1$
by the normalization of the wave function.
The symbol $\otimes$ represents vector addition of the cluster spins and
the orbital angular momentum ${\bf L}$.
We choose the lowest negative-parity excitation with $\ell = 1$.
For $j_{s\bar s}=1$, $j_c$ (${\bf J}_c = {\bf S}_{uud} + {\bf L}$)
can either be 1/2 or 3/2 because $s_{uud}=1/2$ and $\ell=1$,
As in Ref. \cite{TYO97}, we assume that the two possible states have the
same amplitude.
We also limit our consideration to color-singlet cluster configurations,
assuming that hidden color configurations do not contribute to the
single (one-step) knockout processes \cite{HKW92,TYO97}.
Our analyses show that the different $s\bar s$ configurations play
different role in the knockout production.

When the incoming photon strikes the $s \bar s$ cluster, we have
the $s \bar s$-knockout process as shown in Fig. \ref{fig:KO}(a), and
Fig. \ref{fig:KO}(b) corresponds to the $uud$-knockout.
In the $s\bar s$-knockout process, symmetry property of the spatial wave
functions in the initial proton state only allows for magnetic
transition to contribute, while electric (spin-independent)
transition is forbidden.
Then the transition amplitude is proportional to the matrix element
\begin{eqnarray}
\langle S_\phi=1 \vert \mbox{\boldmath ${\sigma}$}_s
-\mbox{\boldmath ${\sigma}$}_{\bar s}
\vert j_{\bar s s} =0,1 \rangle
\cdot \, \left( {\bf q} \times {\bbox{\varepsilon}}_\gamma \right),
\end{eqnarray}
so that only the antisymmetric initial state with
$j_{s\bar s} =0$ contributes.
This leads to $T_{fi}^{s\bar s} \propto b_0$.
In the case of $uud$ knockout, the $s\bar s$-cluster is a spectator, and
only $ j_{\bar s s} =1$ state can match the physical outgoing $\phi$ meson.
Here, both the electric and magnetic transitions contribute and
$T_{fi}^{uud} \propto b_1.$

The detailed description of the knockout process with the relativistic
harmonic oscillator quark model and its electromagnetic current can be
found in Refs. \cite{TOY94,TYO97}.
In this paper we just quote the relevant results.
The knockout amplitudes are most easily evaluated in the laboratory
frame as given in Ref. \cite{TYO97}.
After transforming into the c.m. frame, they read
\begin{eqnarray}
T^{s\bar s }_{m_\phi,m_f;\lambda_\gamma,m_i} &=& i  T^{s\bar s}_0
{\cal S}^{s \bar s}_{m_\phi,m_f;\lambda_\gamma,m_i},
\nonumber \\
T^{uud}_{m_\phi,m_f;\lambda_\gamma,m_i} &=& i T^{uud}_0
{\cal S}^{uud}_{m_\phi,m_f;\lambda_\gamma,m_i}.
\label{TKO}
\end{eqnarray}
Here $T_0^{s\bar s }$ and $T^{uud}_0$ include the dependence of the
amplitudes on the energy and momentum transfer, and ${\cal S}^{s \bar s}$
and ${\cal S}^{uud}$ contain their spin structure.
Explicitly they take the form,
\begin{eqnarray}
T_0^{s \bar s} &=&
\left( \frac{8\pi\alpha_e E_\phi^L E_{p'}^L }{M_N}\right)^{1/2}
A^* b_0 F_{s \bar s}(\gamma_\phi^L, q_{s \bar s})
F_{uud} (\gamma_{p'}^L, 0) V_{s \bar s} ({\bf p}_L')
\frac{\mu_s E_\gamma^L}{3 M_N},
\nonumber \\
T_0^{uud} &=& \left( \frac{8\pi\alpha_e E_\phi^L E_{p'}^L }{M_N}
\right)^{1/2} A^* b_1 F_{s \bar s} (\gamma_\phi^L,0)
F_{uud} (\gamma_{p'}^L, q_{uud}) V_{uud}({\bf q}_L)
\frac{\mu E_\gamma^L}{2M_N},
\end{eqnarray}
and
\begin{eqnarray}
{\cal S}_{fi}^{s\bar s} &=& \sqrt3 \sum_\varrho
\langle {\textstyle\frac12}\, m_f \,
1\, \varrho\, | \, {\textstyle\frac12} \,
m_i \rangle \xi^{s\bar s}_\varrho \lambda_\gamma \bbox{\varepsilon}_\phi^*
(m_\phi) \cdot \bbox{\varepsilon}_\gamma (\lambda_\gamma),
\nonumber \\
{\cal S}_{fi}^{uud} &=&
-\sqrt3 \sum_{\varrho,j_c,m_c} \langle {\textstyle\frac12}\,
m_f-\lambda_\gamma\, 1\, \varrho | j_c\, m_c \rangle\, \langle j_c\, m_c\,
1\, m_\phi | {\textstyle\frac12}\, m_i \rangle\, \xi_\varrho^{uud},
\end{eqnarray}
where
\begin{eqnarray}
  && \xi^{s \bar s}_{\pm 1} = \pm
  \frac{1}{\sqrt2} \sin\theta_{p'},
  \quad \xi^{uud}_{\pm 1} = \mp
  \frac{1}{\sqrt2} \sin\theta_{q},
\nonumber \\
  && \xi^{s \bar s}_0 = \cos\theta_{p'},
  \quad \xi^{uud}_0 = \cos\theta_{q},
\end{eqnarray}
with $\theta_\alpha$ being the production angle in the laboratory frame.
In addition, we use $\mu_s = M_N / M_s$, and $\mu = M_N / M_q$, with
$s$ quark mass $M_s$ ($=500$ MeV) and $u$,$d$ quark mass $M_q$ ($=330$ MeV).
The functions $F_\beta$'s ($\beta = s \bar s, uud$) are the Fourier
transforms of the overlap of the spatial wave functions of the struck
cluster $\beta$ in the entrance and exit channels \cite{TYO97}, which read
\begin{eqnarray}
F_{s\bar s} (\gamma_\phi^L, q_{s\bar s}) &=&
(\gamma_\phi^L)^{-1} \exp ( -r_{s\bar s}^2 q_{s\bar s}^2 /6)
= (\gamma_\phi^L)^{-1} \exp \{-q_{s\bar s}^2/(8\Omega_\rho)\},
\nonumber \\
F_{uud} (\gamma_{p'}^L, q_{uud}) &=&
(\gamma_{p'}^L)^{-2} \exp ( - r_{uud}^2
q_{uud}^2 / 6) = (\gamma_{p'}^L)^{-2} \exp \{ -q_{uud}^2/(6\Omega_\xi)\},
\label{F-I}
\end{eqnarray}
with
\begin{eqnarray}
&& \gamma_{p'}^L = \frac{E_{p'}^L}{M_N}, \qquad
q_{uud}^2 = 2(E_\gamma^L)^2
 - \frac{E_\gamma^L}{E_{p'}^L} [ (E_\gamma^L)^2 + {\bf p}^{\prime 2}_L -
{\bf q}_L^2 ], \nonumber \\
&& \gamma_\phi^L = \frac{E_\phi^L}{M_\phi}, \qquad
q_{s\bar s}^2 = 2(E_\gamma^L)^2 - \frac{E_\gamma^L}{E_{p'}^L}
[ (E_\gamma^L)^2 - {\bf p}^{\prime 2}_L + {\bf q}_L^2 ],
\end{eqnarray}
where $r_{uud}$ and $r_{s\bar s}$ are the rms radii of the proton and
$\phi$ meson, respectively, and $\Omega_{\rho,\xi}$ are the harmonic
oscillator parameters.
We use the parameters determined in Ref. \cite{TYO97} as
$\sqrt{\Omega_\xi} = 1.89$ fm$^{-1}$ and $\sqrt{\Omega_\rho} = 3.02$
fm$^{-1}$.

The momentum distribution function $V_\beta (p)$ of cluster $\beta$ is
given by
\begin{eqnarray}
&& \frac{1}{(2\pi)^3} V_\beta ({\bf p}) =
\frac{v_\beta ({\bf p})}
{\int d{\bf p} \, v_{\beta} ({\bf p})},
\nonumber \\
&& v_\beta ({\bf p}) = {\bf p}^2
\exp \left\{ -\frac{5}{3\Omega_\chi}
\left( {\bf p}^2 - x_\beta M_N E_\beta \right) \right\},
\label{V-I}
\end{eqnarray}
where $x_{s\bar s} = 3/5$, $E_{s\bar s} = E_{p'}^L$ and $x_{uud} = 2/5$,
$E_{uud} = E_\phi^L$.
The parameter $\Omega_\chi$ is again related to the hadron rms radii and
taken to be $\sqrt{\Omega_\chi} = 2.63$ fm$^{-1}$ \cite{TYO97}.

Note that all knockout amplitudes are purely imaginary, which indicates the
absorption of incoming photon by the 5-quark component of the proton.
Therefore, they do not interfere with the OPE amplitude in the
differential cross section.
However, we do expect a strong interference between the dominant imaginary
part of the VDM photoproduction and knockout amplitudes.

%%%%%%%%%%%%%%%%%%%% Results %%%%%%%%%%%%%%%%%%%%%%%

\section{Results}

It is straightforward, with the help of Eq. (\ref{hel}), to obtain
the helicity amplitudes of the knockout and OPE processes.
The total photoproduction helicity amplitude $H$ is given by
\begin{equation}
H = H^{\rm VDM} + H^{s\bar s} + H^{uud} +H^{\rm OPE}.
\end{equation}
We can then proceed to calculate various spin observables with
the formulas developed in Sec. II and Appendix B.
Among those presented in Appendix B, we focus on those which are found
to be strongly dependent on the strangeness content of the proton.

\subsection{Unpolarized cross section}

Before studying the spin observables, let us discuss the parameters
of our model.
In addition to the parameters of the VDM and RHOQM fixed in Sec. III,
we have to determine the amplitudes $b_{0,1}$ of the proton wave
function (\ref{protonwf}).
As we will see, the prediction on the spin observables is sensitive
to the combination $A^*b_{j_{s\bar s}} \equiv \eta_{j_{s\bar s}}
|A^*b_{j_{s\bar s}}|$, where $\eta_{j_{s\bar s}}$ ($= \pm 1$) is the
relative phase between the strange and non-strange amplitudes.
In principle, the purpose of this study is to determine these values
by comparing the predictions with the experimental data.
However, because of the lack of experimental data, we will make an
assumption about these values and compare our results with the pure
VDM and OPE predictions that are associated with the $B^2=0$ case.
For simplicity, we assume $b_0^2 = b_1^2 = B^2/2$.

\begin{figure}
\centerline{\epsfig{file=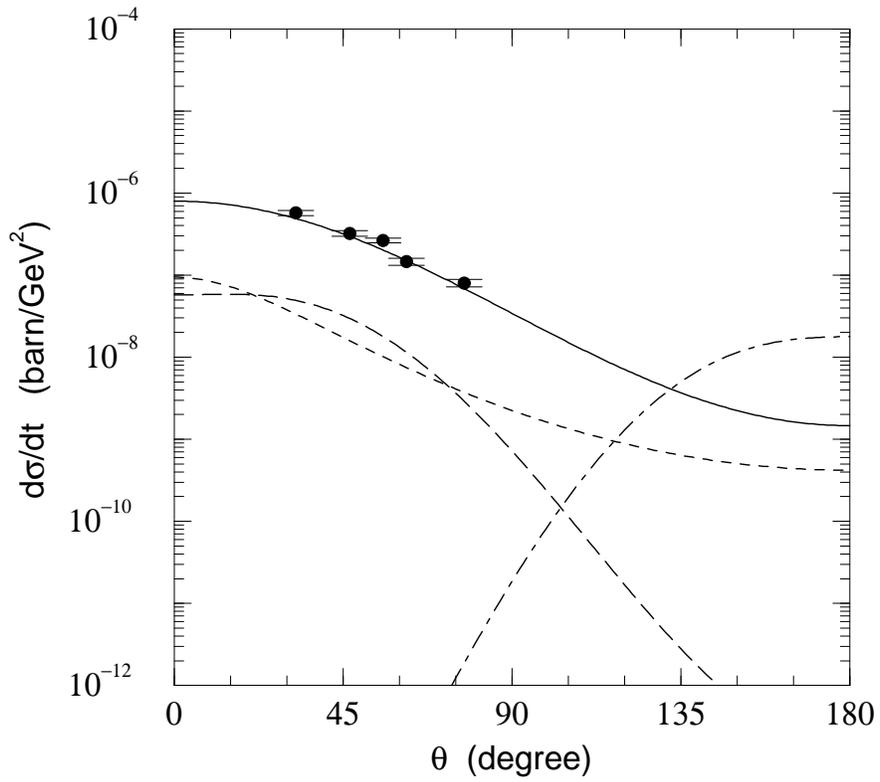, width=0.7\hsize}}
\caption{The unpolarized photoproduction cross section
$d\sigma/dt (\theta)$ at $W = 2.155$ GeV ($E_\gamma^L = 2.0$ GeV).
The solid, dotted, dashed, and dot-dashed lines give the cross section
of VDM, OPE, $s \bar s$-knockout, and $uud$-knockout, respectively,
with strangeness admixture $B^2 = 1$\% and $|b_0| = |b_1| = B^2 /
\protect\sqrt{2}$.
The experimental data are from Ref. \protect\cite{BHKK74}.}
\label{fig:DCS}
\end{figure}

The result of our numerical calculation on the unpolarized $\phi$
photoproduction within the RHOQM is shown in Fig. \ref{fig:DCS}.
In Ref. \cite{TYO97}, we have argued that a theoretical upper-bound of
$B^2$ would be around 3--5\%.
In Fig. \ref{fig:DCS}, we carry out the calculation with the
strangeness probability $B^2=0.01$.
We find that the VDM process dominates the knockout and OPE mechanisms
except in the backward scattering region.
However, our results at large scattering angles should not be
taken seriously because, in this region the applicability of the VDM is
questionable and the contributions from the intermediate excited hadronic
states are expected to be important.
Therefore, the VDM gives the dominant contribution to the cross section
in the kinematical region at small scattering angles in which we are
interested in.

\subsection{Polarization observables}

\begin{figure}
\centerline{\epsfig{file=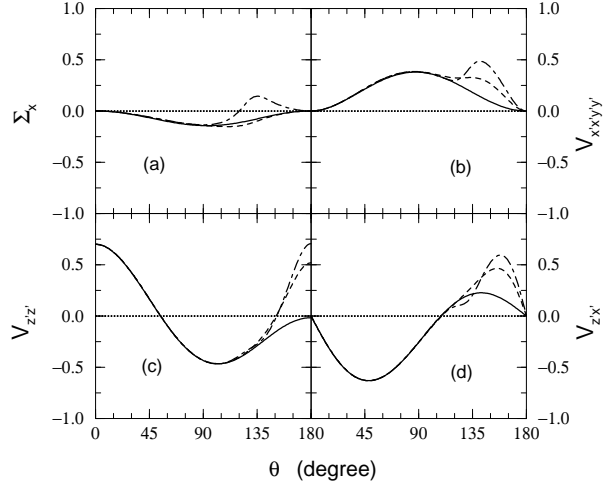, width=8cm}}
\bigskip
\caption{
The single spin observables (a) $\Sigma_x$, (b) $V_{x'x'y'y'}$,
(c) $V_{z'z'}$, and (d) $V_{z'x'}$ at $W = 2.155$ GeV
($E_\gamma^L = 2.0$ GeV).
The solid lines represent the predictions of VDM plus OPE, i.e., $B^2=0$,
and the dashed and dot-dashed lines are for $(\eta_0=+1,\eta_1=+1)$ and
$(+1,-1)$ with $B^2 = 1$\%, respectively. The dependence on $\eta_0$ is
negligible and the results for $(-1,+1)$ and $(-1,-1)$ are not shown.}
\label{fig:SNGL}
\end{figure}

We show our predictions for the single polarization asymmetries,
$\Sigma_x$, $V_{x'x'y'y'}$, $V_{z'z'}$, and $V_{z'x'}$,
in Fig. \ref{fig:SNGL}.
It turns out that the single polarization asymmetries are not
sensitive to the strange quark admixture of the proton.
However, the story is totally different for the double polarization
asymmetries, namely, some of them are very sensitive to the strange
admixture in the proton.
Before presenting our numerical results for double polarization
observables, we first discuss qualitatively why they are important.

Let us consider the most interesting region of $t$, i.e.,
$|t| \to |t|_{\rm min}$ (or $\theta \to 0$), where the differential
cross section is maximal.
Here we can neglect the $uud$-knockout mechanism because the $uud$-knockout
cross section is suppressed in the forward scattering region.
As shown in Eq. (\ref{hel-VDM1}), the diffractive photoproduction
amplitude has the following helicity conserving form in this region
\begin{eqnarray}
H^{\rm VDM}_{\lambda_\phi, \lambda_f;\lambda_\gamma,\lambda_i}
\simeq -i M_0^{\rm VDM}\,
\delta_{\lambda_f\, \lambda_i }\delta_{\lambda_\phi\, \lambda_\gamma},
\label{HVDM}
\end{eqnarray}
where $M_0^{\rm VDM}$ denotes the corresponding amplitude at
$|t| \sim |t|_{\rm min}$.

For the $s\bar s$-knockout amplitude we use
${\hat p'}_\nu \simeq \delta_{\nu\,0}$ at $|t| \simeq |t|_{\rm min}$
and $ \langle  {\textstyle\frac{1}{2}}\, \lambda_i \, 1\, 0 \vert
{\textstyle\frac{1}{2}}\, \lambda_i\, \rangle = 2\lambda_i/\sqrt{3}$
to obtain
\begin{eqnarray}
H^{s\bar s\,}_{\lambda_\phi, \lambda_f;\lambda_\gamma,\lambda_i} \simeq
-iM^{s\bar s\,}_0\, (2\lambda_i\lambda_\gamma)
\,\delta_{\lambda_f\, \lambda_i }\delta_{\lambda_\phi\, \lambda_\gamma}.
\label{Hss}
\end{eqnarray}
Comparison of the helicity dependence of Eqs. (\ref{HVDM}) and (\ref{Hss})
shows that the $s\bar s$-knockout helicity conserving amplitude has an
additional important phase factor $(2\lambda_i\lambda_\gamma)$.
Here, the $\lambda_\gamma$ factor comes from the magnetic structure of
the electromagnetic interaction while $2 \lambda_i$ results from the
coupling of ${\bf S}_{uud}$ with ${\bf L}$ in the initial proton.
The OPE amplitude in this region reads
\begin{eqnarray}
H^{\rm OPE}_{\lambda_\phi, \lambda_f;\lambda_\gamma,\lambda_i} \simeq
- M^{\rm OPE}_0\, (2\lambda_i\lambda_\gamma)
\,\delta_{\lambda_f\, \lambda_i } \delta_{\lambda_\phi\, \lambda_\gamma}.
\end{eqnarray}
Then the total photoproduction amplitude at small $\theta$ becomes
\begin{eqnarray}
H_{\lambda_\phi, \lambda_f;\lambda_\gamma,\lambda_i} \simeq
\left[ -i\left(M^{\rm VDM}_0
 +2\lambda_i\lambda_\gamma M^{s\bar s}_0\right)
 -2\lambda_i\lambda_\gamma M^{\rm OPE}_0\right]
\,\delta_{\lambda_f\, \lambda_i } \delta_{\lambda_\phi\, \lambda_\gamma}.
\end{eqnarray}
Note that in most calculations the Pomeron exchange
amplitude is assumed to be almost imaginary by the optical theorem.
In this approximation, the OPE amplitude does not interfere with
the rest because all the other amplitudes are purely imaginary.
However, the VDM amplitude may have some real part that could interfere
with the OPE contribution \cite{TO97}.
One may estimate this part by using the subtracted dispersion relation
for the amplitude $f(s,t)$, which is normalized to
$s\sigma_T= \mbox{Im}\, f(s,t_{\rm max})$ with $s=W^2$ \cite{Bronzan},
\begin{eqnarray}
\mbox{Re}\, f(s,t)=\frac{2s^2}{\pi}\mbox{P}\int_{s_{\rm min}}^\infty
\frac{ds'}{s'({s'}^2-s^2)}\mbox{Im}\, f(s,t).
\label{dr}
\end{eqnarray}
In Ref. \cite{Bronzan} this integral was evaluated analytically
in the limit of high energy.
Unfortunately, however, this method cannot be applied to the finite $s$
region and we must evaluate Eq. (\ref{dr}) numerically.
Assuming the standard $s$-dependence of the imaginary part as
$f\sim s^{\alpha_P}$ with $\alpha_P \simeq 1$,
we can get the ratio $\xi \equiv \mbox{Re}\, f(s,t)/\mbox{Im}\,
f(s,t)=0.12 \sim 0.086$ at $E_\gamma = 2 \sim 3$ GeV.
Therefore, we are justified to assume the real part of the VDM amplitude
as
\begin{eqnarray}
\mbox{Re}\, H^{\rm VDM}_{\lambda_\phi, \lambda_f,\lambda_\gamma,\lambda_i,}=
- \xi \,M_0^{\rm VDM} \,\delta_{\lambda_f\, \lambda_i }\,
\delta_{\lambda_\phi\, \lambda_\gamma}.
\end{eqnarray}
Then the total amplitude reads
\begin{eqnarray}
H_{\lambda_\phi, \lambda_f,\lambda_\gamma,\lambda_i,}\simeq
-\left[i\left(M^{\rm VDM}_0
+2\lambda_i\lambda_\gamma\,M^{s\bar s}_0\right)
+\left(\xi M^{\rm VDM}_0+2\lambda_i\lambda_\gamma M^{\rm
OPE}_0\right)\right]
\,\delta_{\lambda_f\, \lambda_i } \delta_{\lambda_\phi\, \lambda_\gamma}.
\label{ampl0}
\end{eqnarray}

As an example, let us consider the beam-target asymmetry $C^{\rm
BT}_{zz}$ for the circularly polarized photon beam.
It may be written as
\begin{eqnarray}
C^{\rm BT}_{zz}=
\frac{{d\sigma}(\frac12) -{d\sigma}(\frac32)}
{{d\sigma}(\frac12) +{d\sigma}(\frac32)},
\end{eqnarray}
where $d\sigma$ represents $d\sigma/dt$ and $\frac32$ and $\frac12$ denote
the sum of the initial proton and photon helicities.
In the former case $\lambda_i$ and $\lambda_\gamma$ have the same sign
while in the latter they have opposite signs.
Thus we get
\begin{eqnarray}
C^{\rm BT}_{zz} &\simeq&
\frac
{\Bigl\vert i(M^{\rm VDM}_0-M^{s\bar s}_0)+ \xi M^{\rm VDM}_0-M^{\rm
OPE}_0\Bigl\vert^2
-\Bigl\vert i(M^{\rm VDM}_0+M^{s\bar s}_0)+ \xi M^{\rm VDM}_0+M^{\rm
OPE}_0\Bigl\vert^2}
{\Bigl\vert i(M^{\rm VDM}_0-M^{s\bar s}_0)+ \xi M^{\rm VDM}_0-M^{\rm
OPE}_0\Bigl\vert^2
+\Bigl\vert i(M^{\rm VDM}_0+M^{s\bar s}_0)+ \xi M^{\rm VDM}_0+M^{\rm
OPE}_0\Bigl\vert^2}
\nonumber\\
&\simeq&
-2\,\frac
{({M^{\rm VDM}_0}M^{s\bar s}_0) + \xi M_0^{\rm VDM} M_0^{\rm OPE}}
{|M^{\rm VDM}_0|^2}
\nonumber\\
&\simeq&
-2\,\eta_0 \sqrt{ \frac{\sigma^{s\bar s}}{\sigma^{\rm VDM}} }
-2\, \xi \sqrt{\frac{\sigma^{\rm OPE}}{\sigma^{\rm VDM}} }.
\label{CBT00}
\end{eqnarray}
The above equation explicitly demonstrates the effect of the
$s\bar s$-admixture and the OPE process in the asymmetry.
With the strangeness probability $B^2 = 1$\%, the $s\bar s$-knockout
contribution to the total unpolarized cross section is only at the level
of 5\%.
But in the asymmetry $C^{\rm BT}_{zz}$, its contribution may be seen at
the level of $0.45$ since it is proportional to the square root of the
hidden strangeness contribution to the cross section.
This should be compared with the prediction of VDM plus OPE, which gives
$C^{\rm BT}_{zz} \approx 0$ when the VDM amplitude is purely imaginary.
The OPE contribution to the unpolarized cross section has the same order
of magnitude as that of the $s\bar s$ knockout, and its contribution to
$C^{\rm BT}_{zz}$ comes only from the interference with the real part
of the VDM amplitude.
However, this contribution is suppressed by the additional factor
$\xi \sim 0.1$ and, as a result, it is at the level of 0.05 which
is much smaller than the effect of hidden strangeness in the proton.
Thus, in the results presented below we do not take into account
the real part of the VDM amplitude.

\subsection{Numerical results}

\begin{figure}
\centering
\epsfig{file=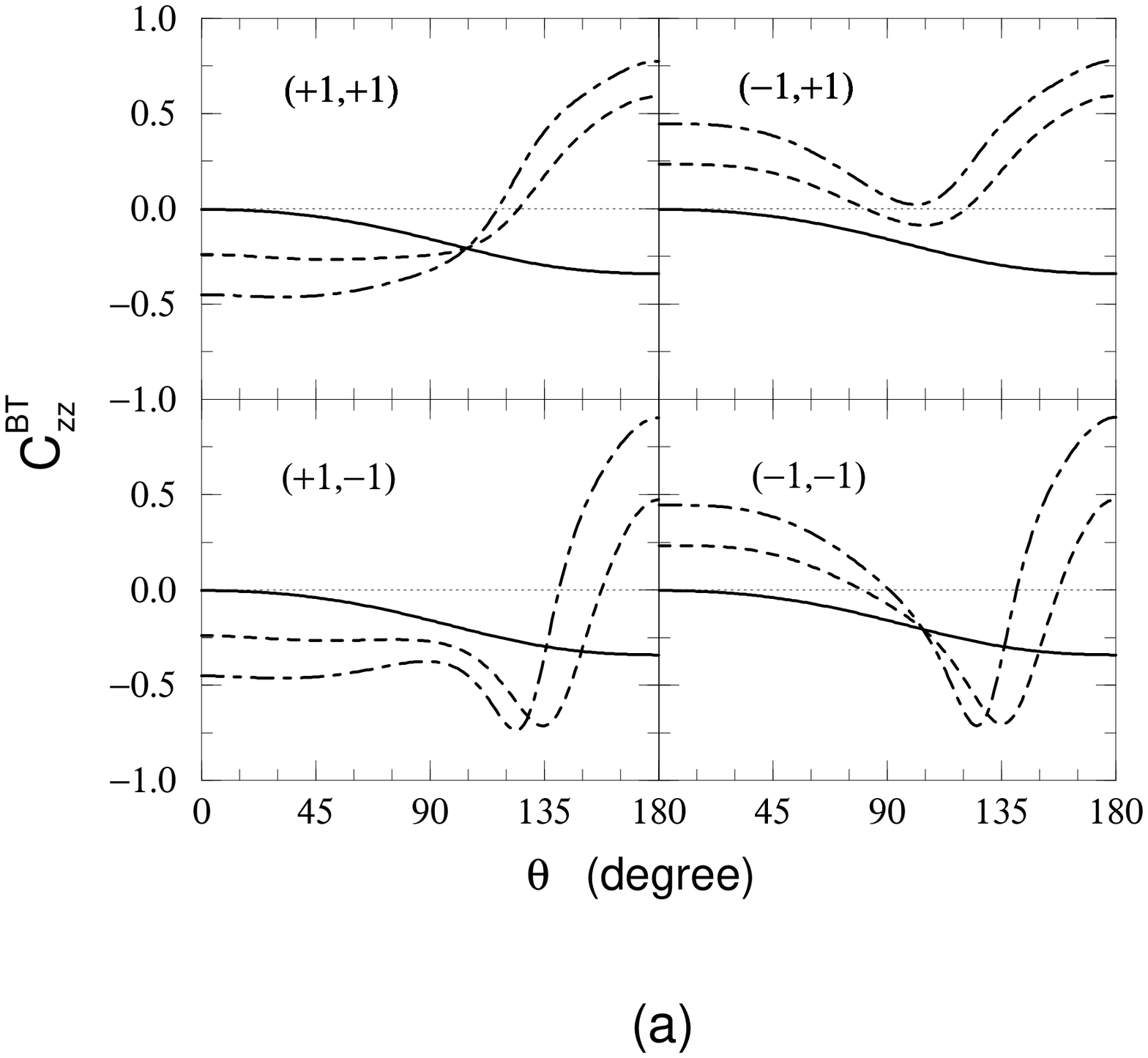, width=7cm} \qquad \qquad
\epsfig{file=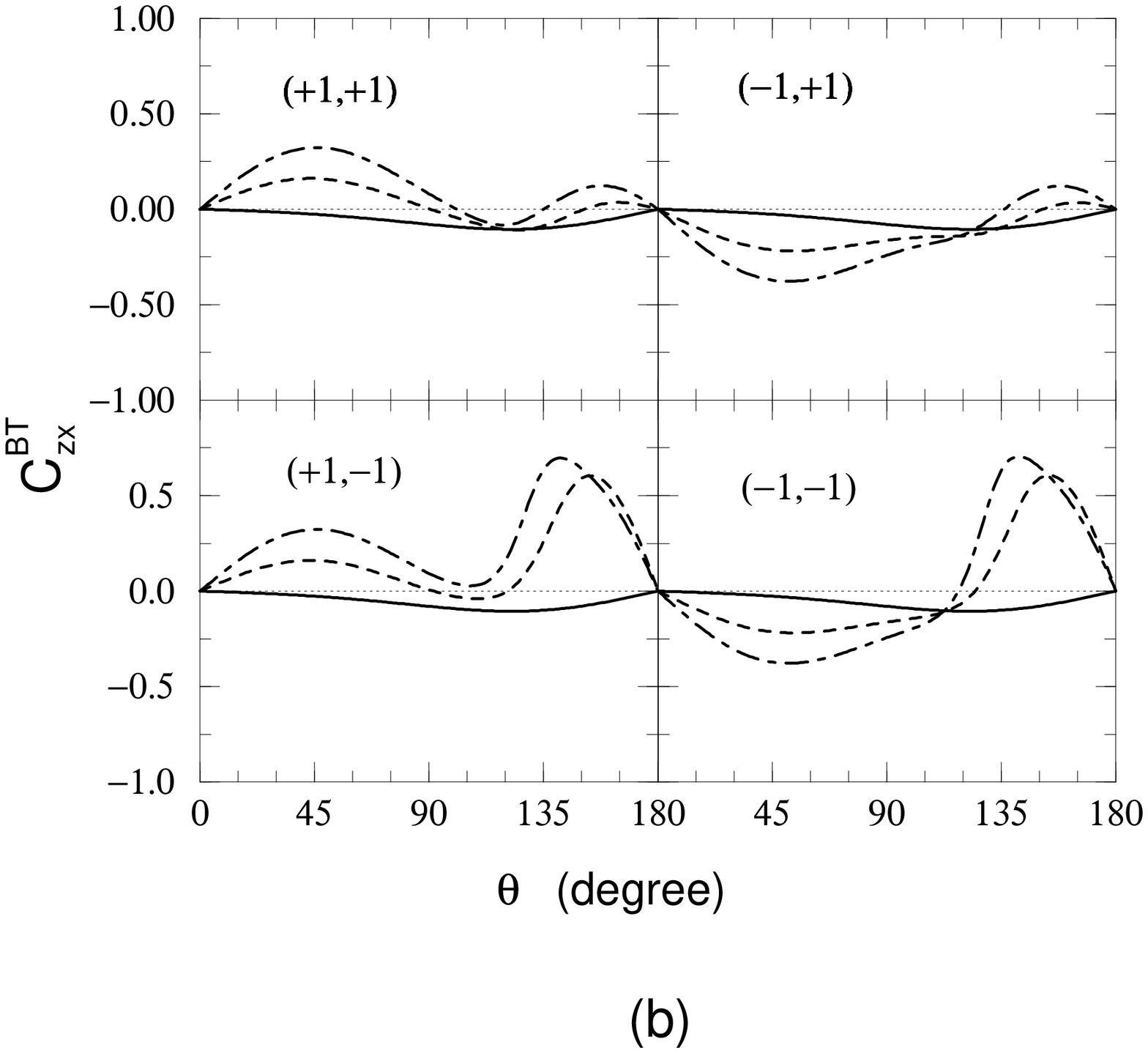, width=7cm}
\caption{The double spin asymmetry (a) $C^{\rm BT}_{zz} (\theta)$ 
and (b) $C^{\rm BT}_{zx} (\theta)$ at
$W = 2.155$ GeV with $B^2=0$\%, i.e., the VDM and OPE (solid lines),
0.25\% (dashed lines), and 1\% (dot-dashed lines) assuming
that $|b_0| = |b_1|$.
The phases $(\eta_0, \eta_1)$ are explicitly given in each graph.}
\label{fig:BT}
\end{figure}

Our results for the Beam--Target double asymmetry%
\footnote{Note that our $C^{\rm BT}_{zz}$ corresponds to the
minus of ${\cal L}_{\rm BT}$ of Ref. \cite{TOY97}.},
$C^{\rm BT}_{ij}$, are shown in Fig. \ref{fig:BT}.
Here the solid line corresponds to the VDM plus OPE prediction,
and the dashed and the dot-dashed lines are the predictions when
we include the knockout contributions with $B^2 = 0.25$\% and $1$\%,
respectively.
Since we have no a priori information about the phases $\eta_{0,1}$,
we give results for all the four different choices of relative phases.
Our numerical calculation confirms the previous qualitative
considerations.
One can see that $C^{\rm BT}_{zz}$ in Fig. \ref{fig:BT}(a) depends
strongly on the hidden strangeness content of the proton
{\em even in the forward scattering region}.
This difference is caused by the different spin structures of the VDM and
the knockout amplitudes.
Therefore, this observable can be used to extract the hidden strangeness
of the proton even for $B^2 \le 1$\%.
The results for $C^{\rm BT}_{zx}$ in Fig. \ref{fig:BT}(b) lead to the same
conclusion although it is not as sensitive as in $C^{\rm BT}_{zz}$.
Note that the results at small $\theta$ are nearly independent
of $\eta_1$.
This is because the $uud$-knockout process is suppressed compared
with other mechanisms in this region.
Similarly, the results are nearly independent of the phase $\eta_0$ at
large $\theta$.
{}From the energy dependence of the polarization observables,
we observe that the knockout contribution is suppressed at
higher energies because of the strong suppression due to the form factors
in the knockout amplitudes.
This leads to the conclusion that the optimal range of the initial
photon energy needed to measure the $s \bar s$ component of the proton
would be around 2--3 GeV.
Furthermore, we find that the forward scattering region of
$\theta \le 30^\circ$ offers a better opportunity to measure the
hidden strangeness contribution.
This conclusion holds for the other spin observables as will be seen below.

In Fig. \ref{fig:BR}, we give our results for the Beam--Recoil
asymmetries, $C^{\rm BR}_{zx'}$ and $C^{\rm BR}_{zz'}$.
This shows that these observables can be useful in probing the
strangeness of the nucleon.

\begin{figure}
\centering
\epsfig{file=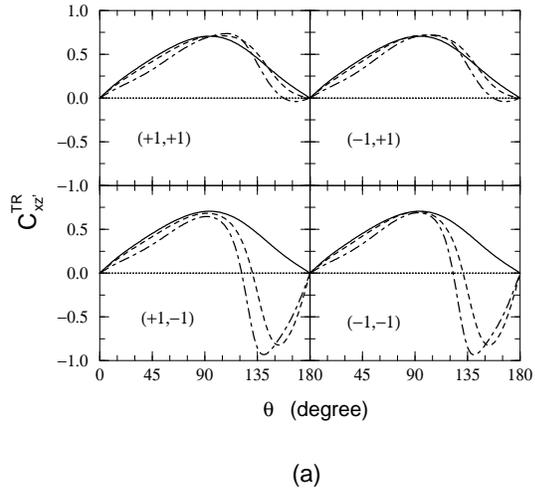, width=7cm} \qquad\qquad
\epsfig{file=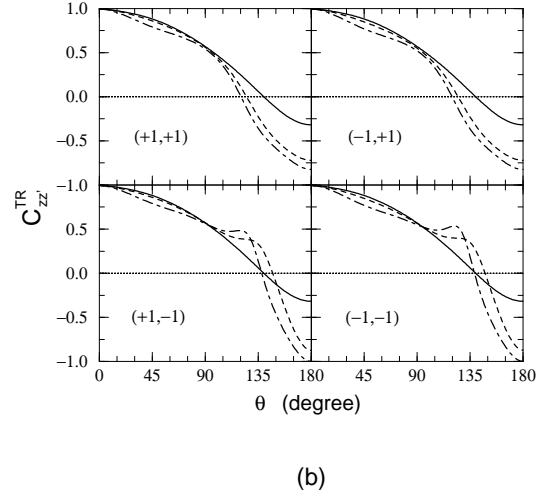, width=7cm}
\caption{Notation same as in Fig. \protect\ref{fig:BT} but for
(a) $C^{\rm TR}_{xz'}$ and (b) $C^{\rm TR}_{zz'}$.}
\label{fig:TR}
\end{figure}

\begin{figure}
\centering
\epsfig{file=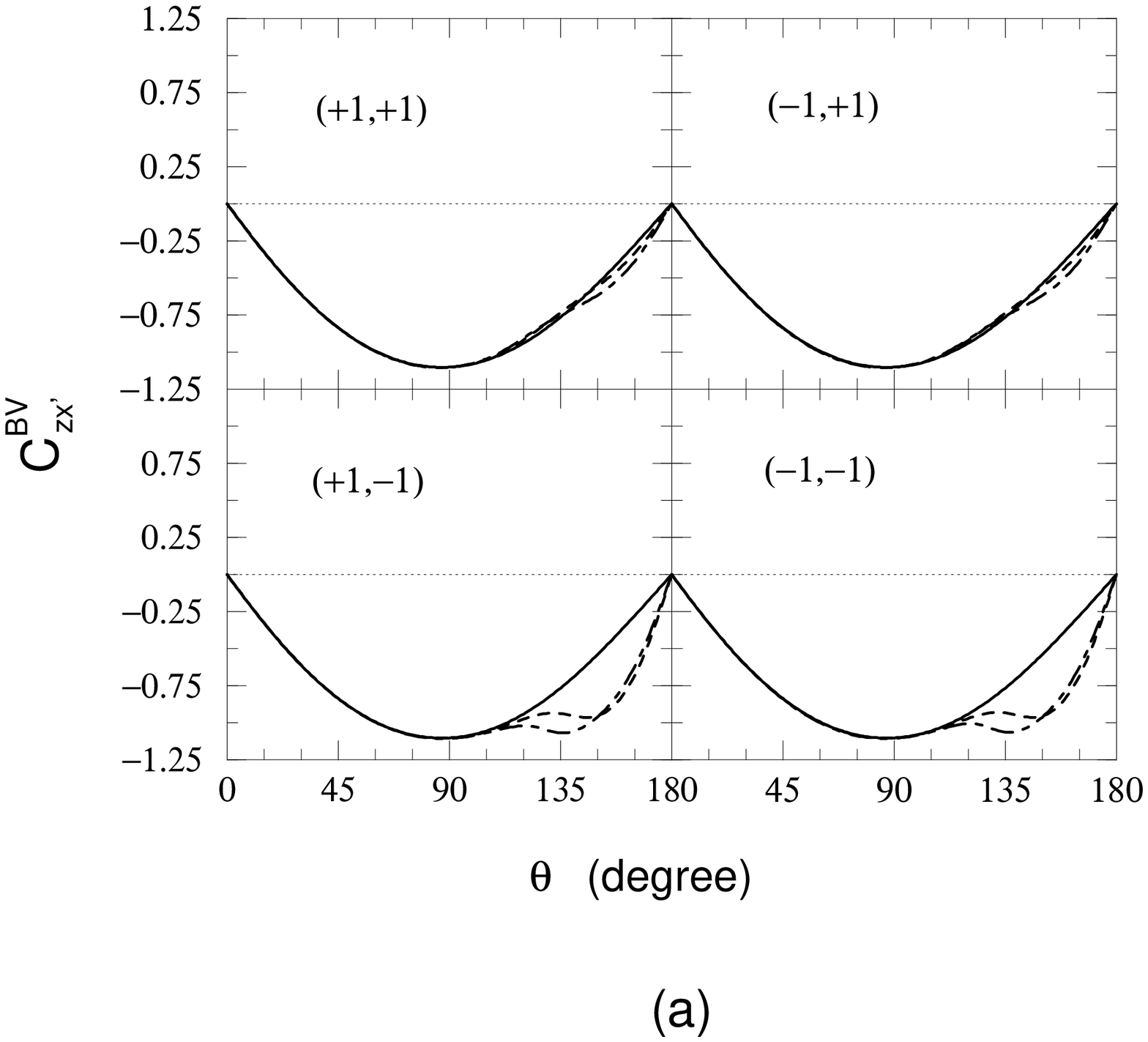, width=7cm} \qquad\qquad
\epsfig{file=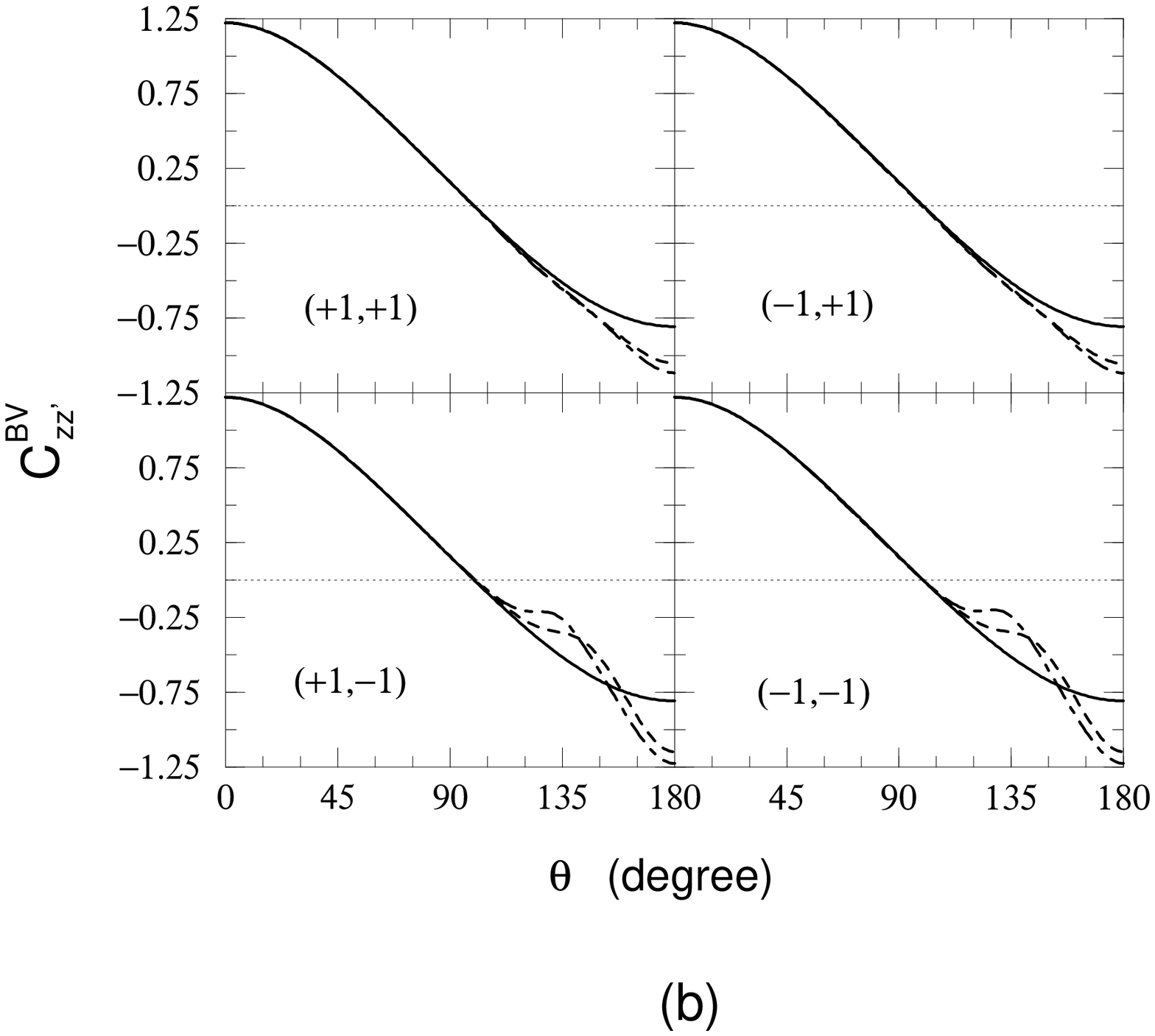, width=7cm}
\caption{Notation same as in Fig. \protect\ref{fig:BT} but for
(a) $C^{\rm BV}_{zx'}$ and (b) $C^{\rm BV}_{zz'}$.}
\label{fig:BV}
\end{figure}

The Target--Recoil double asymmetries $C^{\rm TR}_{xz'}$ and
$C^{\rm TR}_{zz'}$ are shown in Fig. \ref{fig:TR}.
In this case, however, the knockout mechanism gives a very similar
behavior of VDM except at large angles.
Thus the observables $C^{\rm TR}_{xz'}$ and $C^{\rm TR}_{zz'}$ are
{\it not\/} so useful for the purpose of extracting the knockout process.
The same conclusion applies to the Beam--Vector-meson asymmetries
$C^{\rm BV}_{ij}$.
As an example, we give our results in Fig. \ref{fig:BV}, which shows
that $C^{\rm BV}_{zx'}$ and $C^{\rm BV}_{zz'}$ are nearly independent
of the hidden strangeness content of the proton.%
\footnote{Note that the quantities $C^{\rm BV}_{ij}$, $C^{\rm TV}_{ij}$, and
$C^{\rm RV}_{ij}$ with $(j=x',y',z')$ are defined to vary between
$\pm \sqrt{3/2}$ \cite{Ohl72}.}

\begin{figure}
\centering
\epsfig{file=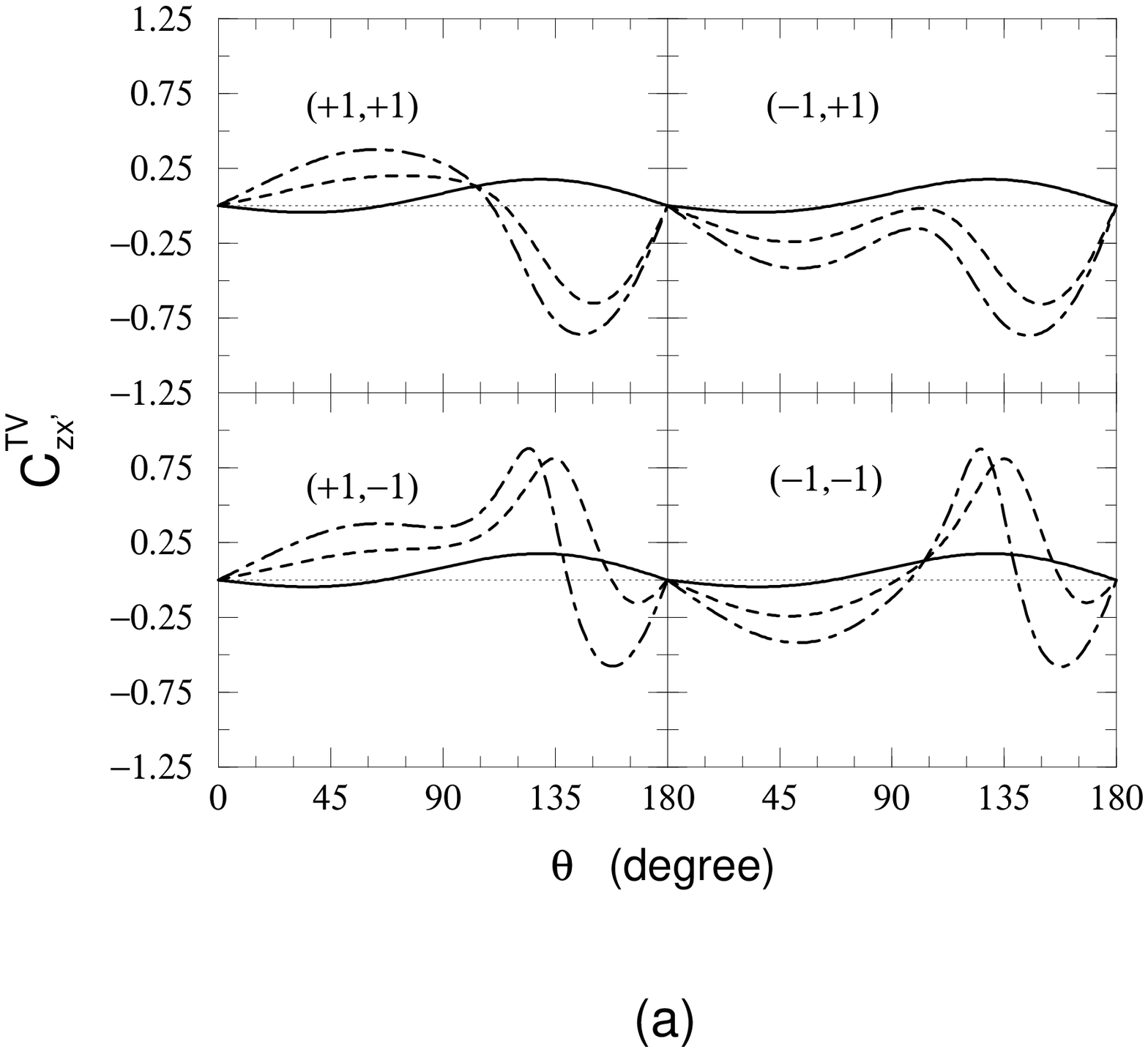, width=7cm} \qquad\qquad
\epsfig{file=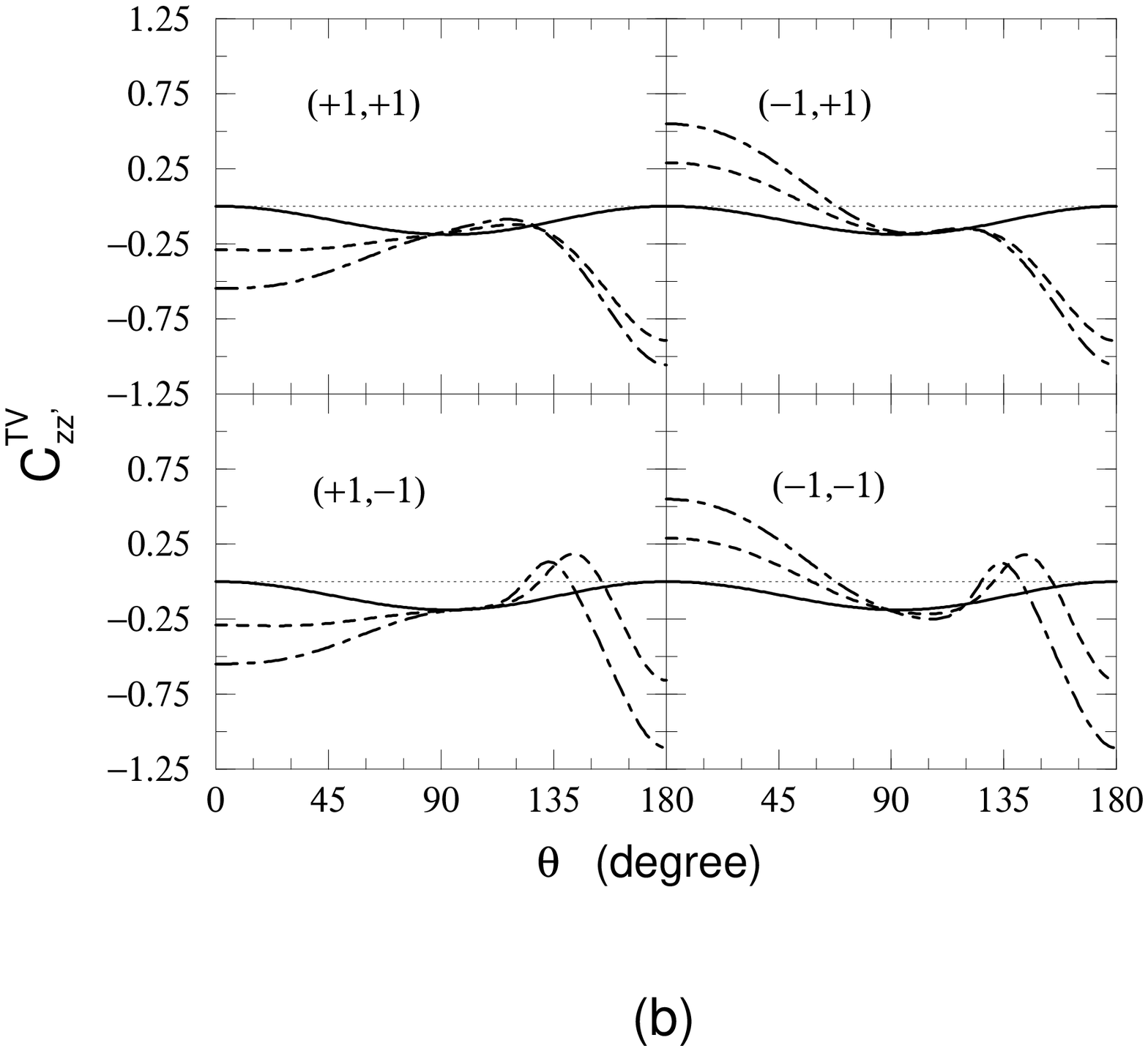, width=7cm}
\caption{Notation same as in Fig. \protect\ref{fig:BT} but for
(a) $C^{\rm TV}_{zx'}$ and (b) $C^{\rm TV}_{zz'}$.}
\label{fig:TV}
\end{figure}

\begin{figure}
\centering
\epsfig{file=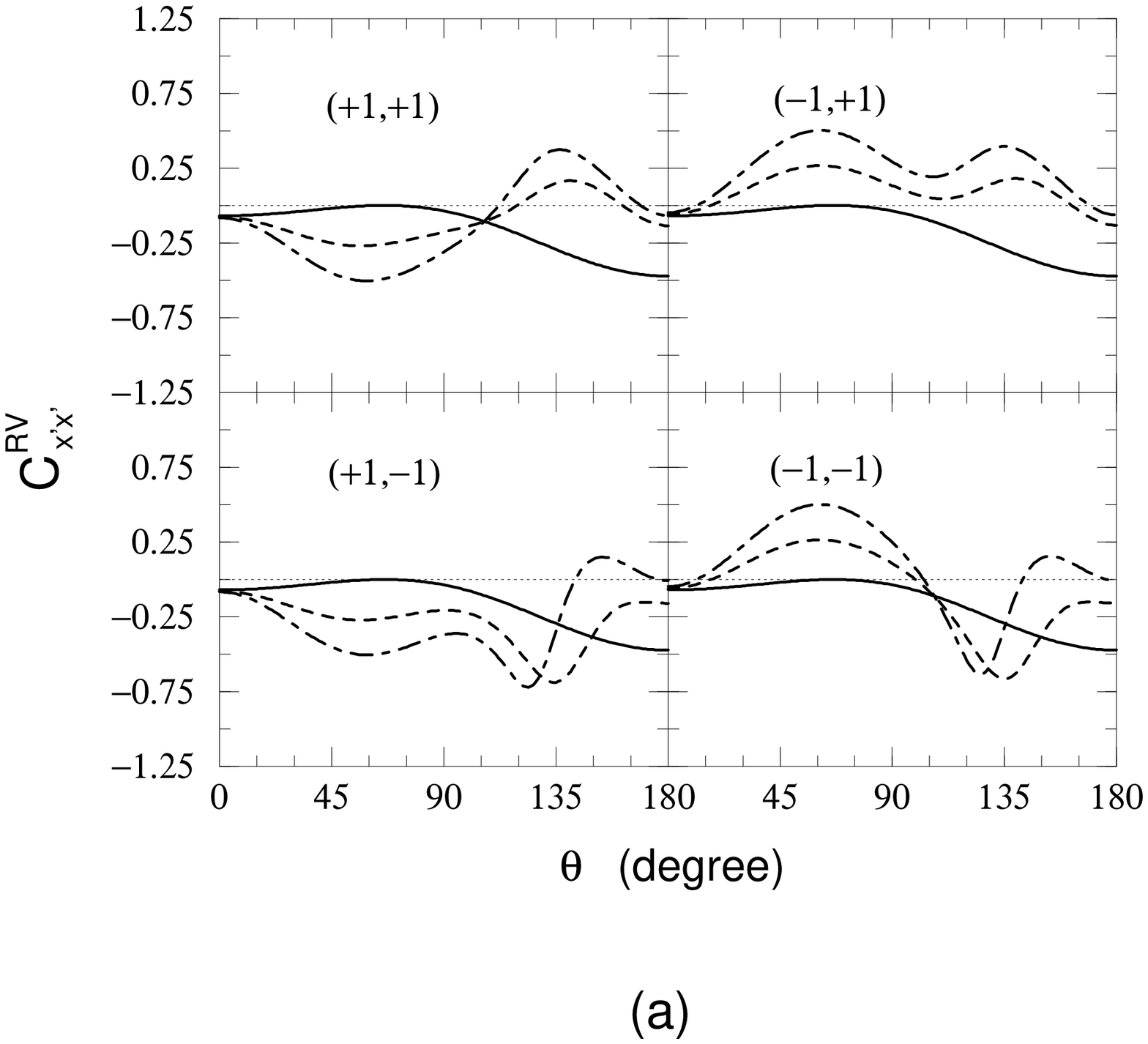, width=7cm} \qquad\qquad
\epsfig{file=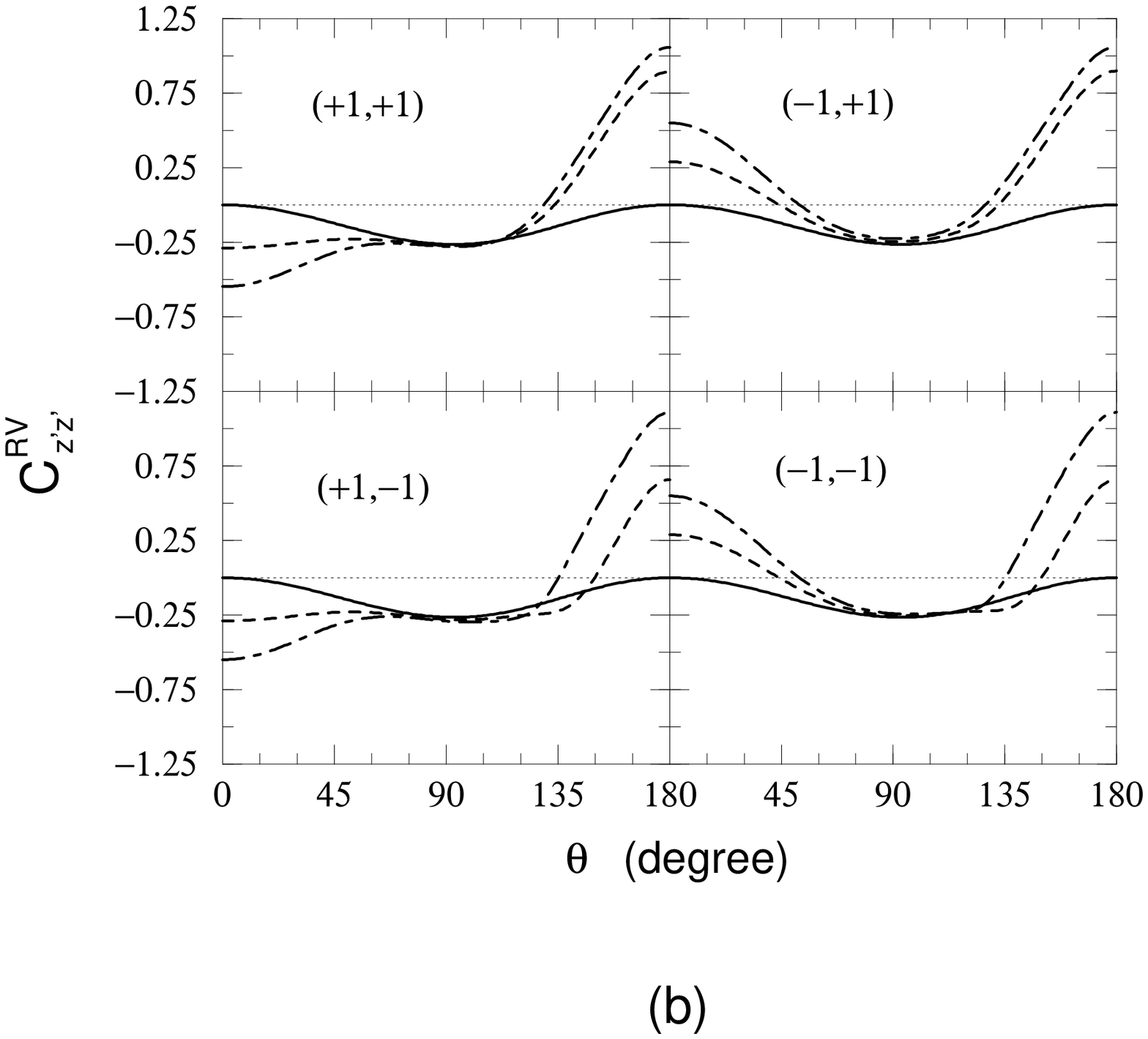, width=7cm}
\caption{Notation same as in Fig. \protect\ref{fig:BT} but for
(a) $C^{\rm RV}_{x'x'}$ and (b) $C^{\rm RV}_{z'z'}$.}
\label{fig:RV}
\end{figure}

Figure \ref{fig:TV} shows our results for $C^{\rm TV}$.
We also present the predictions for $C^{\rm RV}$ in Fig. \ref{fig:RV}.
We see that all of them show strong sensitivity to the strangeness
content of the proton.

%%%%%%%%%%%%%%%%%%%%%  Others  %%%%%%%%%%%%%%%%%%%%%%%%%%%

\section{Other configurations of nucleon wave function}

In the last Section, we assume that both the $s \bar s$ and $uud$
clusters are in their lowest orbital configuration, namely, $S$-state.
We label this configuration as ``Configuration (I).''
In this Section, we discuss the role of the orbitally excited cluster
configurations in the 5-quark cluster model for the nucleon in $\phi$
photoproduction.
We consider the orbital excitation of the $s \bar s$ cluster,
called ``Configuration (II)'' and the orbitally excited $uud$-cluster,
called ``Configuration (III).''
In these cases, the $s \bar s$- and $uud$-clusters form a positive
parity physical proton with $\ell=0$.
Then we can generalize the proton wave function as
\begin{eqnarray}
|p\rangle = A | [uud]^{1/2} \rangle +
\sum_{{n = {\rm I,II,III}} \atop {j_{s \bar s}=0,1}} b_{j_{s\bar s}}^{(n)}
| \bbox{[} [uud]^{j_n} \otimes [s\bar s]^{j_{s \bar s}} \otimes
[{\bf L}]\bbox{]}^{1/2} \rangle ,
\label{protonwfg}
\end{eqnarray}
where the superscripts $j_n$ and $j_{s\bar s}$ denote the spin of each
cluster and $|b_0^{(n)}|^2$ and $|b_1^{(n)}|^2$ correspond to the
spin-0 and spin-1 amplitudes of the $s\bar s$ cluster of ``Configuration
$(n)$,'' respectively.
$B^2$, the strangeness admixture of the proton, is then $\sum (b_{j_{s\bar
s}}^{(n)})^2$.
The amplitudes are constrained to be
$A^2 + B^2 = A^2+\sum (b^{(n)}_{j_{s\bar s}})^2 = 1$ by the
normalization of the wave function.

The symmetry properties of the wave functions in the initial and final
states lead to the selection rules for different $s\bar s$
configurations as summarized in Table \ref{tab1}.
We find that from six possible terms of the proton wave function
(\ref{protonwfg}), only four can contribute to the direct
knockout process: two in $s\bar s$-knockout and two in $uud$-knockout.
The other two amplitudes do not contribute to the direct knockout process,
although they can give a contribution to the total hidden strangeness
probability.

By analyzing the amplitudes one can find that the electric transition is
suppressed by the magnetic as in the case of Configuration (I)
\cite{TYO97}.
For example, the suppression factor for Configuration (II) reads
\begin{eqnarray}
f_{\rm (II)} = \frac{|{\bf p}_L'| \sin\theta_{p'}}{E_\gamma^L},
\end{eqnarray}
which can be expressed in the invariant form as
\begin{eqnarray}
f_{\rm (II)}^2 =
-2 M_N^2 \frac{W^2t^2+\left[ W^2\left(W^2-2M_{N}^2-M_\phi^2 \right)
+M_{N}^2\left( M^2_{N}-M^2_{\phi}\right)t \right] + M^4_\phi M^2_N}
{\left( W^2-M_N^2\right)^4}.
\end{eqnarray}
Numerical estimation shows that with $W \sim 2.1$ GeV the suppression
factor $f_{\rm (II)}^2$ reaches its maximum value around
$0.02$ at $t \sim -0.8$ GeV$^2$ and it becomes negligibly
small as $|t| \to |t|_{\rm max}$ or $|t| \to |t|_{\rm min}$.
A similar suppression factor appears in the electric transition of
$uud$-knockout in Configuration (III).
Thus in the region of $t$ of interest to us, where the knockout mechanism
would be important, the contribution of the electric transitions is
negligible and we will consider the magnetic transitions only.

The amplitudes for Configurations (II) and (III) can be calculated
in a straightforward way using the method of Refs. \cite{TOY94,TYO97}.
The corresponding amplitudes have the form as given in Eq. (\ref{TKO})
by replacing $T_0$ and ${\cal S}$ by
\begin{eqnarray}
T_0^{s \bar s \rm (II)} &=&
\left( \frac{8\pi\alpha_e E_\phi^L E_{p'}^L }{M_N} \right)^{1/2}
A^* b^{\rm (II)}_0 F_{s \bar s}^{\rm (II)}(\gamma_\phi^L, q_{s \bar s})
F_{uud}^{\rm (I)} (\gamma_{p'}^L, 0) \bar{V}_{s \bar s} ({\bf p}_L')
\frac{\mu_s E_\gamma^L}{3 M_N},
\\
{\cal S}_{fi}^{s\bar s \rm (II)} &=&
-\lambda_\gamma\, \sqrt{3}
\sum_{j_c=0,1}
\langle 1 \, m_s\, 1\,  0\vert j_c\,m_s \rangle
\langle {\textstyle \frac{1}{2}}\, m_f\, j_c\, m_s \vert
{\textstyle \frac{1}{2}} m_i \rangle
\langle 1 \, m_s\, 1 \, \lambda_i \vert 1\, m_\phi \rangle,
\label{T-II}
\end{eqnarray}
for ``Configuration (II)'' and
\begin{eqnarray}
T_0^{uud \rm (III)} &=&
\left( \frac{8\pi\alpha_e E_\phi^L E_{p'}^L }{M_N} \right)^{1/2}
A^* b_1^{\rm (III)} F_{s \bar s}^{\rm (I)} (\gamma_\phi^L,0)
F_{uud}^{\rm (III)} (\gamma_{p'}^L, q_{uud}) \bar{V}_{uud} ({\bf q}_L)
\frac{\mu E_\gamma^L}{2M_N},
\\
{\cal S}_{fi}^{uud \rm (III)} &=&
-\sqrt{\frac32}
\sum \limits_{j_c=1/2, 3/2}
\langle j_c\, m_f-\lambda_\gamma\, 1\, m_\phi \vert
 {\textstyle\frac{1}{2}}\, m_i \rangle
\langle {\textstyle \frac{1}{2}}\, m_f-\lambda_\gamma\, 1\,0 \vert
j_c\, m_f-\lambda_\gamma \rangle,
\label{T-III}
\end{eqnarray}
for ``Configuration (III).''
The functions $F_\beta^{\rm (II), (III)}$ with
$\beta = (s \bar s, uud)$ are related to the correspondent
$F_\beta^{\rm (I)}$ of Eq. (\ref{F-I}) as
\begin{eqnarray}
F_{s\bar s}^{\rm (II)} (\gamma_\phi^L, q_{s\bar s}) &=&
\frac{E_\gamma^L \sqrt{r^2_{s\bar s}}(1-V_{q\parallel})}{3}\,
F_{s\bar s}^{\rm (I)} (\gamma_\phi^L, q_{s\bar s}),
\nonumber\\
F_{uud}^{\rm (III)} (\gamma_{p'}^L, q_{uud}) &=&
\frac{E_\gamma^L \sqrt{r^2_{uud}}(1-V_{p'\parallel})}{3}\,
F_{uud}^{\rm (I)} (\gamma_{p'}^L, q_{uud}),
\end{eqnarray}
where $V_{p\,\parallel} = |{\bf p}| \cos\theta_{p} /E_{p}^L$.
The momentum distribution function $\bar{V}_\beta (p)$ of cluster $\beta$
is given by
\begin{eqnarray}
\frac{1}{(2\pi)^3} \bar{V}_\beta ({\bf p}) &=&
\frac{\bar{v}_\beta ({\bf p})}{\int d{\bf p} \, v_{\beta}^{(0)} ({\bf p})},
\nonumber \\
\bar{v}_\beta ({\bf p}) &=&
\exp \left\{ -\frac{5}{3\Omega_\chi}
\left( {\bf p}^2 - x_\beta M_N E_\beta \right) \right\}.
\end{eqnarray}
Note that the difference with $v_\beta$ of Eq. (\ref{V-I}) lies in the
absence of the factor ${\bf p}^2$.
The calculation of $T^{s\bar s \rm (II)}$ is rather similar to that
of $T^{s\bar s \rm (I)}$ which is given in Ref. \cite{TYO97} and
Appendix E contains the derivation of $T^{s\bar s \rm (III)}$ in some
detail.

\begin{figure}
\centering
\epsfig{file=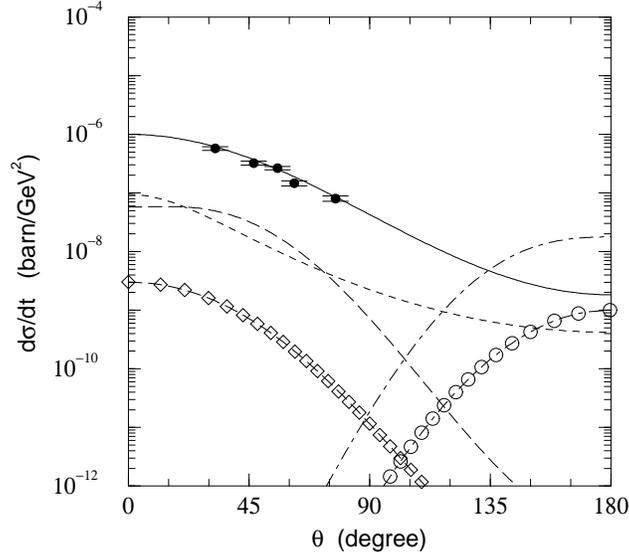, width=0.5\hsize}
\caption{The unpolarized photoproduction cross section
$d\sigma/dt (\theta)$ at $W = 2.155$ GeV ($E_\gamma^L = 2.0$ GeV).
The solid and dotted lines give the cross section of VDM and OPE.
The dashed and dot-dashed lines are from the $s \bar s$- and $uud$-knockout
from Configuration (I) in the proton wave function.
The dashed line with diamonds and the dot-dashed line with circles are the
cross sections from the $s \bar s$- and $uud$-knockout by assuming only
Configurations (II) and (III) in the nucleon, respectively.
The strangeness admixture is assumed to be $B^2 = 1$\% and
$|b_0^{(n)}| = |b_1^{(n)}| = B^2 / \protect\sqrt{2}$.
The experimental data are from Ref. \protect\cite{BHKK74}.}
\label{fig:DCS-F}
\end{figure}

Analyses of the relative contribution from different cluster
configurations show that the contribution of Configurations (II) and
(III) are much smaller (by an order of magnitude) than that of
Configuration (I) even if we assume the same values for $b_{0,1}^{(n)}$.
This can be seen from Fig. \ref{fig:DCS-F} where we present our results
for the differential cross sections from each configuration.

For clarity, let us consider the $s\bar s$-knockout for
Configurations (I) and (II).
The ratio of the spatial matrix elements for Configurations (I) and
(II) read
\begin{eqnarray}
R_{s\bar s}=
\frac{E_\gamma^L (1-V_{q\,\parallel})\sqrt{r^2_{s\bar s}}}{|{\bf p}'_L|}
\frac{N_{\rm II}}{3N_{\rm I}},
\end{eqnarray}
where $N_{\rm I,II}$ are the normalization factors of the radial
wave functions of Configurations (I) and (II), respectively.
Since
\begin{eqnarray}
N^{-2}_{\alpha} \sim\int d{\bf p}v_\alpha({\bf p}),
\end{eqnarray}
we can obtain
\begin{eqnarray}
\frac{N_{\rm II}}{N_{\rm I}}\simeq
\sqrt{\frac{3\Omega_\chi}{2}}.
\end{eqnarray}
Using the numerical value of the dimensional parameters
$r_{s\bar s}=0.29$ fm and $\sqrt{\Omega_\chi}=2.63$ fm$^{-1}$
\cite{TYO97}, we obtain
\begin{eqnarray}
R_{s\bar s}\simeq 0.31
\frac{E_\gamma^L (1-V_{q\,\parallel})  }{|{\bf p}'_L|}.
\end{eqnarray}
Since at $\theta \simeq 0$ we have $\cos\theta_{p'} \simeq \cos\theta_\phi
\simeq 1$, $E_\phi^L = E_\gamma^L + t/2M_N \simeq E_\gamma^L$, and
\begin{eqnarray}
|{\bf p}'_L| = |{\bf k}_L| - |{\bf q}_L| \simeq E_\gamma^L
\left( 1-\frac{|{\bf q}_L|}{E_\phi^L} \right) =
E_\gamma^L (1-V_{q\,\parallel}),
\end{eqnarray}
we then obtain $R_{s\bar s}^2 \simeq 0.1$, which agrees with the
numerical calculation of Fig. \ref{fig:DCS-F}.
A similar conclusion can be drawn for the $uud$-knockout from
Configurations (I) and (III).

The above analysis shows that the cross section of the knockout process is
dominated by Configuration (I) and we can safely neglect the other
cluster configurations in the proton wave function.
Now let us consider the polarization observables.
For simplicity, we again consider the case of Configuration (II).
From the amplitude (\ref{T-II}), one can find
\begin{equation}
H^{s\bar s \rm (II)}_{\lambda_f,\lambda_\phi;\lambda_\gamma,\lambda_i}
\propto \delta_{\lambda_f\, \lambda_i} \,
\delta_{\lambda_\phi\, \lambda_\gamma}.
\end{equation}
This has the same structure as the VDM helicity amplitude (\ref{HVDM}).
Since its amplitude is suppressed by the dominant VDM amplitude, however,
it cannot be extracted from the background VDM contribution.
We could verify this analysis by numerical calculation even with
the assumption of the same values for $b_{j_{s\bar s}}^{(n)}$.
Furthermore, because of their heavy mass, the coefficients $b_{0,1}$ of
Configurations (II) and (III) are expected to be
much smaller than those of Configuration (I).
As a conclusion, therefore, the contributions from the orbitally excited
cluster configurations can be neglected in the polarization observables
as well.

%%%%%%%%%%%%%%%%%%%%%  Summary  %%%%%%%%%%%%%%%%%%%%%%%%%%%

\section{Summary and Conclusion}

We have studied the possibility of using the spin observables of
the $\phi$ meson photoproduction process in probing the hidden
strangeness content of the proton.
We consider the direct knockout mechanism in addition to the VDM
and OPE processes by assuming an $s\bar s$ component in the proton
wave function.
Unlike the differential cross section, we find that the spin
observables could be useful in disentangling the knockout process from
the VDM and OPE processes.
We find that single polarization observables are {\em not\/} sensitive
to the strangeness content of the proton.
However, some double polarization observables, notably,
$C^{\rm BT}_{zx,zz}$, $C^{\rm BR}_{zx',zz'}$, $C^{\rm TV}_{zx',zz'}$,
and $C^{\rm RV}_{xx',zz'}$, are {\it very sensitive\/} to the hidden
strangeness content of the proton in the forward scattering region,
whereas most of the Target--Recoil and Beam--Vector-meson double
asymmetries are not.
It indicates that measurements of these double polarization observables
could be very useful in probing the strangeness content of the proton.

We also find that the contribution of the knockout mechanism is
suppressed with increasing initial photon energy because of the
strong suppression due to the form factors in the knockout amplitudes.
Therefore, we expect that the optimal range of the initial photon energy
needed to measure the $s \bar s$ component of the proton would be
around 2--3 GeV.
However, it should be mentioned that at extremely low energy just near
the threshold one has to take into account the OZI evading re-scattering
process and it would be interesting to
study its effect on the polarization observables.

The orbitally excited quark cluster configurations in the proton wave
function was also investigated in connection with $\phi$ photoproduction.
We find that their role is {\em not\/} important in the cross section and
polarization observables and, therefore, these configurations can be
neglected in the study of $\phi$ photoproduction.

The purpose of this study is to determine the strangeness content of
the proton by investigating the polarization observables.
Unfortunately, because of the scarcity of presently available
experimental data \cite{MMSS71,Lowry97}, we can not give any definite
predictions for the strangeness content of the proton based on our analyses.
Thus, new experiments are strongly called for at the current electron
facilities which, hopefully, will help to shed light on our understanding
of the proton structure.

Finally, we point out that, since the $s\bar s$ knockout process
dominates the $uud$ knockout at the forward scattering angle, one
can estimate the value of $b_0$ in the proton wave function
(\ref{protonwf}) by analyzing polarization observables.
However, it is not easy to get estimate for $b_1$ because its
contribution can be seen only at large $\theta$ where
corrections to our model are expected to be important.
To get information for $b_1$, therefore, it would be interesting to
apply our analyses to $\eta (\eta')$ photoproduction as a complementary
process to $\phi$ photoproduction, since the $s\bar s$ knockout process
in this case is associated with $b_1$.

%%%%%%%%%%%%%%%%%  Acknowledgment  %%%%%%%%%%%%%%%%%%%%%%%

\acknowledgements

We gratefully acknowledge the useful discussions with M. Fujiwara,
S. B. Gerasimov, S. V. Goloskokov, C. R. Ji, T. Kinashi, and M. Namiki.
Y.O. is also grateful to D.-P. Min for encouragement and wish to thank
the Physics Department and the Center for Theoretical Sciences of the
National Taiwan University for the warm hospitality.
A.I.T. appreciates the warm hospitality of the Faculty of
Human Development of Kobe University where part of this work was
carried out.
This work was supported in part by the Russian Foundation for Basic
Research under grant No. 96-15-96423,
the Korea Science and Engineering Foundation through the Center for
Theoretical Physics of Seoul National University,
the National Science Council of ROC under grant No. NSC87-2112-M-002,
and Monbusho's Special Program for Promoting Advanced
Study (1996, Japan).
Lastly we thank A. Jackson of Kobe Shoin Women's College for
careful reading of the manuscript.

%%%%%%%%%%%%%%%%%%%%%  Appendix  %%%%%%%%%%%%%%%%%%%%%%%%%%

\appendix

\section{Density matrices}

In this Appendix, we discuss the density matrices of the photon, target
and recoil proton, and the vector meson.
In general, the density matrix of the photon can be written as
\begin{equation}
\rho_\gamma = \frac12 ( \bbox{1}_2 + \bbox{\sigma}_\gamma \cdot
{\bf P}_S),
\end{equation}
in photon helicity space, where $\bbox{1}_2$ is the $2 \times 2$ unit
matrix and ${\bf P}_S$ is the Stokes vector which defines the direction
and degree of polarization of the photon beam.
The presence of $\bbox{\sigma}_\gamma$ is due to the fact that a real
photon has only two spin degrees of freedom.
The Stokes vectors corresponding to some special cases of photon
polarization can be found, for example, in Ref.  \cite{FTS92}.

The proton density matrix is in the spin $\frac12$ space and is therefore
a $2 \times 2$ Hermitian matrix. So we have
\begin{equation}
\rho_N = \frac12 ( \bbox{1}_2 + \bbox{\sigma}_N \cdot {\bf P}_N ),
\end{equation}
for the target proton and
\begin{equation}
\rho_{N'} = \frac12 ( \bbox{1}_2 + \bbox{\sigma}_{N'} \cdot {\bf P}_{N'}
),
\end{equation}
for the recoil proton.

For the vector meson, because of its spin-1 structure, the density matrix
cannot be described by vector polarizations only.
To describe the vector meson polarization completely, we have to take into
account the tensor polarizations. The tensor polarization operator is
defined as \cite{BLS80,Con94}
\begin{equation}
S_{jk} = \frac32(S_j S_k + S_k S_j) - 2 \delta_{j\,k} \bbox{1}_3 ,
\end{equation}
where $\bbox{1}_3$ is the $3 \times 3$ unit matrix with
\begin{equation}
S_x = \frac{1}{\sqrt2} \left( \begin{array}{ccc} 0 & 1 & 0 \\ 1 & 0 & 1 \\
0 & 1 & 0 \end{array} \right), \qquad
S_y = \frac{1}{\sqrt2} \left( \begin{array}{ccc} 0 & -i & 0 \\ i & 0 & -i
\\
0 & i & 0 \end{array} \right), \qquad
S_z = \left( \begin{array}{ccc} 1 & 0 & 0 \\ 0 & 0 & 0 \\
0 & 0 & -1 \end{array} \right).
\end{equation}
Only five of them are independent since $S_{xx} + S_{yy} + S_{zz} = 0$.
Therefore, we are led to the final form of the density matrix of vector
meson as
\begin{equation}
\rho_V = \frac13 (\bbox{1}_3 + \sum_j P^V_j \Omega^V_j ),
\end{equation}
where
\begin{equation}
\Omega^V_j = \sqrt{\frac32}( S_x, S_y, S_z), \quad
\frac{1}{\sqrt6} ( S_{xx}-S_{yy}),  \quad
\frac{1}{\sqrt2} S_{zz}, \quad
\sqrt{\frac23} ( S_{xy}, S_{yz}, S_{zx} ),
\end{equation}
which are normalized as $\mbox{Tr}\, \Omega^V_j \Omega^V_k = 3
\delta_{j\,k}$.

The explicit forms of the matrices appearing in Eq. (\ref{genspin}) are
$(\bbox{1}_2,\bbox{\sigma}_\gamma)$ for $A_\gamma$,
$(\bbox{1}_2,\bbox{\sigma}_{N(N')})$ for $A_N$ ($B_{N'}$), and
$(\bbox{1}_3,\Omega^V_j)$ for $B_V$.%
\footnote{One has to use $-\sigma_x$ and $-\sigma_z$ for the initial and
final protons instead of $\sigma_x$ and $\sigma_z$ in order to have the
correct helicity states in the c.m. system \cite{FTS92}.}

\section{Single and Double Polarization Observables in Helicity Amplitudes}

In this Appendix, we give the explicit expressions for the spin observables
in terms of helicity amplitudes.

The cross section intensity ${\cal I}(\theta)$ is defined as
\begin{equation}
{\cal I}(\theta) = \frac14 \,\mbox{Tr}\, ( {\cal F} {\cal F}^\dagger ),
\end{equation}
which leads to
\begin{equation}
{\cal I}(\theta) = \frac12 \sum_{i=1}^4 \sum_{a=\pm 1,0} |H_{i,a}|^2.
\end{equation}

The explicit expressions for non-vanishing single polarization observables
are as follows.
\begin{mathletters}
\begin{eqnarray}
\Sigma_x \cdot {\cal I}(\theta) &=& -\mbox{Re} \left\{ H_{4,1}^* H_{1,-1} -
H_{4,0}^* H_{1,0} + H_{4,-1}^* H_{1,1} \right. \nonumber \\
&& \hskip 0.5cm \left. \mbox{} - H_{3,1}^* H_{2,-1} + H_{3,0}^* H_{2,0} -
H_{3,-1}^* H_{2,1} \right\}.
\\
T_y \cdot {\cal I}(\theta) &=& -\mbox{Im}\, \left\{ H_{4,-1}^* H_{3,-1} +
H_{4,0}^* H_{3,0} + H_{4,1}^* H_{3,1} \right. \nonumber \\
&& \left. \hskip 0.5cm \mbox{} + H_{2,-1}^* H_{1,-1} + H_{2,0}^* H_{1,0}
+
H_{2,1}^* H_{1,1} \right\}.
\\
P_{y'} \cdot {\cal I}(\theta) &=& - \mbox{Im}\, \left\{ H_{4,-1}^* H_{2,-1}
+ H_{4,0}^* H_{2,0} + H_{4,1}^* H_{2,1}\right. \nonumber \\
&& \left. \hskip 0.5cm \mbox{} + H_{3,-1}^* H_{1,-1} + H_{3,0}^* H_{1,0} +
H_{3,1}^* H_{1,1} \right\}.
\\
V_{y'} \cdot {\cal I}(\theta) &=& - \frac{\sqrt3}{2} \mbox{Im}\,
\left\{ H_{4,0}^* ( H_{4,1} - H_{4,-1} ) + H_{3,0}^* ( H_{3,1} - H_{3,-1})
\right. \nonumber
\\
&& \left. \hskip 0.5cm \mbox{} + H_{2,0}^* (H_{2,1} - H_{2,-1}) + H_{1,0}^*
(H_{1,1} - H_{1,-1}) \right\}.
\\
V_{x'x'y'y'} \cdot {\cal I}(\theta) &=& \sqrt{\frac32} \mbox{Re}
\left\{ H_{4,-1}^* H_{4,1} + H_{3,-1}^* H_{3,1} + H_{2,-1}^* H_{2,1} +
H_{1,-1}^* H_{1,1} \right\}.
\\
V_{z'z'} \cdot {\cal I}(\theta) &=& \frac{1}{2\sqrt2} \left\{ |H_{4,-1}|^2 -
2 |H_{4,0}|^2 + |H_{4,1}|^2 + |H_{3,-1}|^2 - 2 |H_{3,0}|^2 + |H_{3,1}|^2
\right. \nonumber\\
&& \left. \hskip 0.5cm \mbox{} + |H_{2,-1}|^2 - 2 |H_{2,0}|^2 + |H_{2,1}|^2
+ |H_{1,-1}|^2 - 2 |H_{1,0}|^2 + |H_{1,1}|^2 \right\}.
\\
V_{z'x'} \cdot {\cal I}(\theta) &=& \frac{\sqrt3}{2} \mbox{Re}
\left\{ H_{4,0}^* (H_{4,1} - H_{4,-1}) + H_{3,0}^* (H_{3,1} - H_{3,-1})
\right. \nonumber
\\
&& \left. \hskip 0.5cm \mbox{} + H_{2,0}^* (H_{2,1} - H_{2,-1}) + H_{1,0}^*
(H_{1,1} - H_{1,-1}) \right\}.
\end{eqnarray}
\end{mathletters}

The explicit expressions for some double polarization observables are given
below.

\noindent\noindent
$\bullet$ Beam--Target

\begin{mathletters}
\begin{eqnarray}
C^{\rm BT}_{yx} \cdot {\cal I}(\theta) &=&
\mbox{Im}\, \left\{ H_{4,-1}^* H_{2,1} -
H_{4,0}^* H_{2,0} + H_{4,1}^* H_{2,-1} \right. \nonumber \\
&& \left. \hskip 0.5cm \mbox{} - H_{3,-1}^* H_{1,1} + H_{3,0}^*
H_{1,0} - H_{3,1}^* H_{1,-1} \right\},
\\
C^{\rm BT}_{yz} \cdot {\cal I}(\theta) &=&
- \mbox{Im}\, \left\{ H_{4,-1}^* H_{1,1} -
H_{4,0}^* H_{1,0} + H_{4,1}^* H_{1,-1} \right. \nonumber \\
&& \left. \hskip 0.5cm \mbox{} + H_{3,-1}^* H_{2,1} - H_{3,0}^*
H_{2,0} + H_{3,1}^* H_{2,-1} \right\},
\\
C^{\rm BT}_{zx} \cdot {\cal I}(\theta) &=&
- \mbox{Re}\, \left\{ H_{4,-1}^* H_{3,-1} +
H_{4,0}^* H_{3,0} + H_{4,1}^* H_{3,1} \right. \nonumber \\
&& \left. \hskip 0.5cm \mbox{} + H_{2,-1}^* H_{1,-1} + H_{2,0}^* H_{1,0} +
H_{2,1}^* H_{1,1} \right\},
\\
C^{\rm BT}_{zz} \cdot {\cal I}(\theta) &=& - \frac12 \left\{ |H_{4,-1}|^2 +
|H_{4,0}|^2 + |H_{4,1}|^2 - |H_{3,-1}|^2 - |H_{3,0}|^2 - |H_{3,1}|^2
\right. \nonumber \\
&& \left. \mbox{} + |H_{2,-1}|^2 + |H_{2,0}|^2 + |H_{2,1}|^2 -
|H_{1,-1}|^2 - |H_{1,0}|^2 - |H_{1,1}|^2 \right\}.
\end{eqnarray}
\end{mathletters}

\noindent\noindent
$\bullet$ Beam--Recoil

\begin{mathletters}
\begin{eqnarray}
C^{\rm BR}_{yx'} \cdot {\cal I}(\theta) &=&
\mbox{Im}\, \left\{ H_{4,-1}^* H_{3,1} -
H_{4,0}^* H_{3,0} + H_{4,1}^* H_{3,-1} \right. \nonumber \\
&& \left. \hskip 0.5cm \mbox{} - H_{2,-1}^* H_{1,1} + H_{2,0}^* H_{1,0} -
H_{2,1}^* H_{1,-1} \right\},
\\
C^{\rm BR}_{yz'} \cdot {\cal I}(\theta) &=&
\mbox{Im}\, \left\{ H_{4,-1}^* H_{1,1} -
H_{4,0}^* H_{1,0} + H_{4,1}^* H_{1,-1}\right. \nonumber \\
&& \left. \hskip 0.5cm \mbox{} - H_{3,-1}^* H_{2,1} + H_{3,0}^* H_{2,0} -
H_{3,1}^* H_{2,-1} \right\},
\\
C^{\rm BR}_{zx'} \cdot {\cal I}(\theta) &=&
- \mbox{Re}\, \left\{ H_{4,-1}^* H_{2,-1} +
H_{4,0}^* H_{2,0}  + H_{4,1}^* H_{2,1} \right. \nonumber \\
&& \left. \hskip 0.5cm \mbox{} + H_{3,-1}^* H_{1,-1} + H_{3,0}^* H_{1,0} +
H_{3,1}^* H_{1,1} \right\},
\\
C^{\rm BR}_{zz'} \cdot {\cal I}(\theta) &=& \frac12 \left\{ |H_{4,-1}|^2 +
|H_{4,0}|^2 + |H_{4,1}|^2 + |H_{3,-1}|^2 + |H_{3,0}|^2 + |H_{3,1}|^2
\right. \nonumber \\
&& \left. \mbox{} - |H_{2,-1}|^2 - |H_{2,0}|^2 - |H_{2,1}|^2 -
|H_{1,-1}|^2 - |H_{1,0}|^2 - |H_{1,1}|^2 \right\}.
\end{eqnarray}
\end{mathletters}

\noindent\noindent
$\bullet$ Target--Recoil

\begin{mathletters}
\begin{eqnarray}
C^{\rm TR}_{xx'} \cdot {\cal I}(\theta) &=&
\mbox{Re}\, \left\{ H_{4,-1}^* H_{1,-1} +
H_{4,0}^* H_{1,0} + H_{4,1}^* H_{1,1} \right. \nonumber \\
&& \left. \hskip 0.5cm \mbox{} + H_{3,-1}^* H_{2,-1} + H_{3,0}^* H_{2,0} +
H_{3,1}^* H_{2,1} \right\},
\\
C^{\rm TR}_{xz'} \cdot {\cal I}(\theta) &=&
- \mbox{Re}\, \left\{ H_{4,-1}^* H_{3,-1} +
H_{4,0}^* H_{3,0} + H_{4,1}^* H_{3,1} \right. \nonumber \\
&& \left. \hskip 0.5cm \mbox{} - H_{2,-1}^* H_{1,-1} - H_{2,0}^*
H_{1,0} - H_{2,1}^* H_{1,1} \right\},
\\
C^{\rm TR}_{zx'} \cdot {\cal I}(\theta) &=&
\mbox{Re}\, \left\{ H_{4,-1}^* H_{2,-1} +
H_{4,0}^* H_{2,0} + H_{4,1}^* H_{2,1} \right. \nonumber \\
&& \left. \hskip 0.5cm \mbox{}  - H_{3,-1}^* H_{1,-1} - H_{3,0}^* H_{1,0} -
H_{3,1}^* H_{1,1} \right\},
\\
C^{\rm TR}_{zz'} \cdot {\cal I}(\theta) &=& -\frac12 \left\{ |H_{4,-1}|^2 +
|H_{4,0}|^2 + |H_{4,1}|^2 - |H_{3,-1}|^2 - |H_{3,0}|^2 - |H_{3,1}|^2
\right. \nonumber \\
&& \left. \mbox{} - |H_{2,-1}|^2 - |H_{2,0}|^2 - |H_{2,1}|^2 +
|H_{1,-1}|^2 + |H_{1,0}|^2 + |H_{1,1}|^2 \right\}.
\end{eqnarray}
\end{mathletters}

\noindent\noindent
$\bullet$ Beam--Vector-meson

\begin{mathletters}
\begin{eqnarray}
C^{\rm BV}_{yx'} \cdot {\cal I}(\theta) &=&
- \frac{\sqrt3}{2} \mbox{Im}\, \left\{
H_{4,-1}^* H_{1,0} - H_{4,0}^* ( H_{1,-1} + H_{1,1} ) + H_{4,1}^* H_{1,0}
\right. \nonumber \\
&& \left. \mbox{} - H_{3,-1}^* H_{2,0} + H_{3,0}^* ( H_{2,-1} + H_{2,1} ) -
H_{3,1}^* H_{2,0} \right\},
\\
C^{\rm BV}_{yz'} \cdot {\cal I}(\theta) &=&
- \sqrt{\frac32} \mbox{Im}\, \left\{
H_{4,-1}^* H_{1,1}  - H_{4,1}^* H_{1,-1} - H_{3,-1}^* H_{2,1} + H_{3,1}^*
H_{2,-1} \right\},
\\
C^{\rm BV}_{zx'} \cdot {\cal I}(\theta) &=&
\frac{\sqrt3}{2} \mbox{Re}\, \left\{
H_{4,0}^* ( H_{4,-1} + H_{4,1} ) + H_{3,0}^* ( H_{3,-1} + H_{3,1} )
\right. \nonumber \\
&& \left. \hskip 0.5cm \mbox{}
+ H_{2,0}^* ( H_{2,-1} + H_{2,1} ) + H_{1,0}^* ( H_{1,-1} + H_{1,1} )
\right\},
\\
C^{\rm BV}_{zz'} \cdot {\cal I}(\theta) &=&
- \frac12 \sqrt{\frac32} \left\{
|H_{4,-1}|^2 - |H_{4,1}|^2 + |H_{3,-1}|^2 - |H_{3,1}|^2
\right. \nonumber \\
&& \left. \mbox{} + |H_{2,-1}|^2 - |H_{2,1}|^2 + |H_{1,-1}|^2 - |H_{1,1}|^2
\right\}.
\end{eqnarray}
\end{mathletters}

\noindent\noindent
$\bullet$ Target--Vector-meson

\begin{mathletters}
\begin{eqnarray}
C^{\rm TV}_{xx'} \cdot {\cal I}(\theta) &=&
- \frac{\sqrt3}{2} \mbox{Re}\, \left\{
H_{4,0}^* ( H_{3,-1} + H_{3,1} ) + H_{3,0}^* ( H_{4,-1} + H_{4,1} ) \right.
\nonumber \\
&& \left. \hskip 0.5cm \mbox{} + H_{2,0}^* ( H_{1,-1} + H_{1,1} ) +
H_{1,0}^* ( H_{2,-1} + H_{2,1} ) \right\},
\\
C^{\rm TV}_{xz'} \cdot {\cal I}(\theta) &=&
\sqrt{\frac32} \mbox{Re}\, \left\{
H_{4,-1}^*H_{3,-1} - H_{4,1}^* H_{3,1} + H_{2,-1}^* H_{1,-1} - H_{2,1}^*
H_{1,1} \right\},
\\
C^{\rm TV}_{zx'} \cdot {\cal I}(\theta) &=&
- \frac{\sqrt3}{2} \mbox{Re}\, \left\{
H_{4,0}^* ( H_{4,-1} + H_{4,1} ) - H_{3,0}^* ( H_{3,-1} + H_{3,1} ) \right.
\nonumber \\
&& \left. \hskip 0.5cm \mbox{} + H_{2,0}^* ( H_{2,-1} + H_{2,1} ) -
H_{1,0}^* ( H_{1,-1} + H_{1,1} ) \right\},
\\
C^{\rm TV}_{zz'} \cdot {\cal I}(\theta) &=&
\frac12 \sqrt{\frac32} \left\{
|H_{4,-1}|^2 - |H_{4,1}|^2 - |H_{3,-1}|^2 + |H_{3,1}|^2
\right. \nonumber \\
&& \left. \mbox{} + |H_{2,-1}|^2 - |H_{2,1}|^2 - |H_{1,-1}|^2 + |H_{1,1}|^2
\right\}.
\end{eqnarray}
\end{mathletters}

\noindent\noindent
$\bullet$ Recoil--Vector-meson

\begin{mathletters}
\begin{eqnarray}
C^{\rm RV}_{x'x'} \cdot {\cal I}(\theta) &=&
- \frac{\sqrt3}{2} \mbox{Re}\, \left\{
H_{4,0}^* ( H_{2,-1} + H_{2,1} ) + H_{3,0}^* ( H_{1,-1} + H_{1,1} ) \right.
\nonumber \\
&& \left. \hskip 0.5cm \mbox{} + H_{2,0}^* ( H_{4,-1} + H_{4,1} ) +
H_{1,0}^* ( H_{3,-1} + H_{3,1} ) \right\},
\\
C^{\rm RV}_{x'z'} \cdot {\cal I}(\theta) &=&
\sqrt{\frac32} \mbox{Re}\, \left\{
H_{4,-1}^* H_{2,-1} - H_{4,1}^* H_{2,1} + H_{3,-1}^* H_{1,-1} - H_{3,1}^*
H_{1,1} \right\},
\\
C^{\rm RV}_{z'x'} \cdot {\cal I}(\theta) &=&
\frac{\sqrt3}{2} \mbox{Re}\, \left\{
H_{4,0}^* ( H_{4,-1} + H_{4,1} ) + H_{3,0}^* ( H_{3,-1} + H_{3,1} ) \right.
\nonumber \\
&& \left. \hskip 0.5cm \mbox{} - H_{2,0}^* ( H_{2,-1} + H_{2,1} ) -
H_{1,0}^* ( H_{1,-1} + H_{1,1} ) \right\},
\\
C^{\rm RV}_{z'z'} \cdot {\cal I}(\theta) &=&
-\frac12 \sqrt{\frac32} \left\{
|H_{4,-1}|^2 - |H_{4,1}|^2 + |H_{3,-1}|^2 - |H_{3,1}|^2
\right. \nonumber \\
&& \left. \mbox{} - |H_{2,-1}|^2 + |H_{2,1}|^2 - |H_{1,-1}|^2 + |H_{1,1}|^2
\right\}.
\end{eqnarray}
\end{mathletters}

\section{Models on the VDM amplitude}

In Sec. III.A, we write the invariant amplitude of the diffractive
production as
\begin{equation}
T_{fi}^{\rm VDM} = i T_0  \varepsilon^*_{\phi\mu} {\cal M}^{\mu\nu}
\varepsilon_{\gamma\nu},
\end{equation}
with ${\cal M}^{\mu\nu} = {\cal F}_\alpha \Gamma^{\alpha,\mu\nu}$,
where ${\cal F}_\alpha$ and $\Gamma^{\alpha,\mu\nu}$ correspond to
the Pomeron-nucleon vertex and Pomeron--photon--vector-meson vertex,
respectively.

In Ref. \cite{TOY97}, we used
\begin{equation}
\tilde\Gamma^{\alpha,\mu\nu} =
( k + q )^\alpha g^{\mu\nu} - k^\mu \, g^{\alpha\nu} - q^\nu \, g^{\alpha\mu},
\label{Gamold}
\end{equation}
instead of the $\Gamma^{\alpha,\mu\nu}$ of Eq. (\ref{Gamalmunu}).
This expression comes from gauging the massive-vector field Lagrangian
for the $\phi \phi \mathbb{P}$ interaction which is assumed to have the
same spin structure as the $\phi\phi\gamma$ vertex.
However, it does not satisfy the gauge invariance condition
$q_\mu {\cal M}^{\mu\nu} = {\cal M}^{\mu\nu} k_\nu = 0$.
One way to get a gauge invariant amplitude is to multiply
$\tilde\Gamma^{\alpha,\mu\nu}$ by the projection operator
${\cal P}_{\mu\nu}$ as in Sec. III,
\begin{eqnarray}
\tilde\Gamma^{\alpha,\mu\nu} \to \tilde\Gamma^{\alpha,\mu\nu}_1
&=& {\cal P}^{\mu\mu'} \tilde\Gamma^\alpha_{\mu'\nu'} {\cal P}^{\nu'\nu}
\nonumber \\
&=& (k+q)^\alpha g^{\mu\nu} - \frac{(k+q)^\alpha}{k \cdot q} k^\mu q^\nu .
\label{Vold1}
\end{eqnarray}

Another way to project the gauge non-invariant part out of
$\tilde\Gamma^{\alpha,\mu\nu}$ is to multiply ${\cal P}_{(l)}^{\mu\nu}$ and
${\cal P}_{(r)}^{\mu\nu}$ as
\begin{equation}
\tilde\Gamma^{\alpha,\mu\nu} \to \tilde\Gamma^{\alpha,\mu\nu}_2
= {\cal P}_{(l)}^{\mu\mu'} \tilde\Gamma^\alpha_{\mu'\nu'}
{\cal P}_{(r)}^{\nu'\nu},
\end{equation}
where
\begin{eqnarray}
{\cal P}_{(l)}^{\mu\nu} = g^{\mu\nu} - \frac{1}{q^2} q^\mu q^\nu,
\nonumber \\
{\cal P}_{(r)}^{\mu\nu} = g^{\mu\nu} - \frac{1}{k^2} k^\mu k^\nu.
\end{eqnarray}
This gives us
\begin{eqnarray}
\tilde\Gamma^{\alpha,\mu\nu}_2 &=&
 (k+q)^\alpha g^{\mu\nu} - k^\mu g^{\alpha\nu} - q^\nu g^{\alpha\mu}
\nonumber \\ && \mbox{}
-\frac{1}{q^2} q^\mu ( k^\alpha q^\nu - k \cdot q \, g^{\alpha\nu} )
-\frac{1}{k^2} k^\nu ( q^\alpha k^\mu - k \cdot q \, g^{\alpha\mu} ),
\label{Vold2}
\end{eqnarray}
which retains all the terms of $\tilde\Gamma^{\alpha,\mu\nu}$.
In fact, since the $q^\mu$ and $k^\nu$ terms do not contribute after
being multiplied with the boson polarization vectors,
$\tilde\Gamma^{\alpha,\mu\nu}_2$ gives results identical to
that of $\tilde\Gamma^{\alpha,\mu\nu}$ in the calculation.

When we consider the Pomeron exchange model of Sec. III, we obtained
the gauge invariant amplitude (\ref{Gamalmunu}) by applying the projection
operator ${\cal P}_{\mu\nu}$ to the $\bar\Gamma^{\alpha,\mu\nu}$ of
Eq. (\ref{Gam-0}).
One may, however, use the projection operators ${\cal P}_{(l)}^{\mu\nu}$
and ${\cal P}_{(r)}^{\mu\nu}$ to get
\begin{eqnarray}
\bar\Gamma^{\alpha,\mu\nu} \to
\Gamma^{\alpha,\mu\nu}_2
&=& {\cal P}_{(l)}^{\mu\mu'} \bar\Gamma^\alpha_{\mu'\nu'}
{\cal P}_{(r)}^{\nu'\nu} \nonumber \\
&=& (k+q)^\alpha g^{\mu\nu} - 2 k^\mu g^{\alpha\nu}
- 2 q^\nu g^{\alpha\mu} 
\nonumber \\ && \mbox{}
- \frac{k \cdot q}{k^2 q^2} (k+q)^\alpha q^\mu k^\nu 
- \frac{q^\mu}{q^2} \{ k^\alpha q^\nu - q^\alpha q^\nu - 2 k \cdot q \,
  g^{\alpha\nu} \}
\nonumber \\ && \mbox{}
- \frac{k^\nu}{k^2} \{ q^\alpha k^\mu - k^\alpha k^\mu - 2 k \cdot q \,
  g^{\alpha\mu} \},
\label{Gam2}
\end{eqnarray}
where the last three terms vanish after being multiplied with the boson
polarization vectors although they are required to ensure gauge invariance.
Note the close similarity between the amplitudes $\Gamma^{\alpha,\mu\nu}$'s.

\begin{figure}
\centering
\epsfig{file=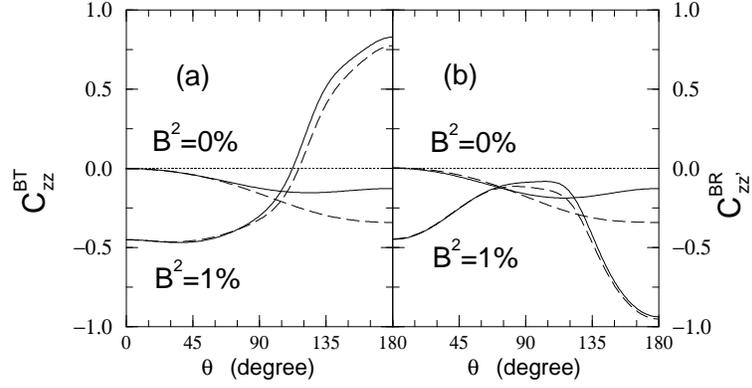, width=0.6\hsize}
\caption{Double polarization asymmetries (a) $C^{\rm BT}_{zz}$ and (b)
$C^{\rm BR}_{zz'}$ for two models
of $\Gamma^{\alpha,\mu\nu}$ with $B^2=0$ and $1$\%.
The solid lines are obtained with $\Gamma^{\alpha,\mu\nu}$ of
(\protect\ref{Gamalmunu}) and the dashed lines with
$\tilde\Gamma^{\alpha,\mu\nu}$ of (\protect\ref{Gamold}).
For simplicity, we take the phases $(\eta_0,\eta_1) = (+,+)$.}
\label{fig:vdmgi}
\end{figure}

Different choice of the gauge invariant $\Gamma^{\alpha,\mu\nu}$'s as given
in Eqs. (\ref{Gamalmunu}), (\ref{Vold1}), (\ref{Vold2}), and (\ref{Gam2})
will necessitate the use of different form for $T_0$ \cite{FS69}.
However, these different forms of $T_0$'s will be related to each other
as they are required to describe the unpolarized cross section.
As it turns out, because of the incompleteness of our model,
the $\Gamma^{\alpha,\mu\nu}$'s of Eqs. (\ref{Vold2}) and (\ref{Gam2})
have a singularity problem in the case of electroproduction as $k^2 \to 0$.
In order to have a model for VDM amplitude which can be applied to
electroproduction, we use the $\Gamma^{\alpha,\mu\nu}$ of Eq.
(\ref{Gamalmunu}) with the $T_0$ of Eq. (\ref{Tform}) in our calculation.
To see the model dependence of our results, we carry out the
calculations for the $\tilde\Gamma^{\alpha,\mu\nu}$ of Eq. (\ref{Gamold})
as well%
\footnote{This is the model adopted in Ref. \cite{TOY97}.}.
Our results for $C^{\rm BT}_{zz}$ in these two models are shown in Fig.
\ref{fig:vdmgi}.
One can see the close similarity of the two model predictions at small
$\theta$.
This is true for the other spin observables as well, namely, they
give nearly the same results in the kinematical region of our interest,
say $\theta \le 30^\circ$.
This is because the dominant contribution in this region comes from
the $(k+q)^\alpha g^{\mu\nu}$ term, which is present in both models
for VDM amplitude.
As a conclusion, the model of this work gives the same predictions with
those of Ref. \cite{TOY97} in the kinematical range of interest in this
study.

\section{Helicity Amplitude in VDM}

In this Appendix, we give the details of the derivation for the helicity
amplitude (\ref{hel-VDM}) of the VDM.
We first define
\begin{equation}
\gamma_p = \frac{E_p^L}{M_N} \qquad \mbox{and} \qquad
\gamma_p' = \frac{E_{p'}^L}{M_N},
\end{equation}
which leads to
\begin{eqnarray}
\alpha &\equiv& \frac{|{\bf p}|}{E_p^L + M_N} =
\sqrt{\frac{\gamma_p-1}{\gamma_p+1}}, \nonumber \\
\alpha' &\equiv& \frac{|{\bf p}'|}{E_{p'}^L + M_N} =
\sqrt{\frac{\gamma_p'-1}{\gamma_p'+1}}.
\end{eqnarray}
Let ${\bf n} = {\bf p} / |{\bf p}|$ and
${\bf n}' = {\bf p}' / |{\bf p}'|$, we can then write
\begin{equation}
u_m(p) = \sqrt{\frac{\gamma_p+1}{2}} \left(
\begin{array}{c} 1 \\ \alpha \bbox{\sigma}\cdot {\bf n} \end{array} \right)
\chi_m,
\end{equation}
for the Dirac spinor of the proton with spin projection $m$,
and a similar expression for $u_{m'}(p')$ for the outgoing proton.

Let us define ${\cal U}^\alpha$ as
\begin{equation}
{\cal U}^\alpha = \varepsilon^*_{\phi\mu} \Gamma^{\alpha,\mu\nu}
\varepsilon_{\gamma\nu}.
\end{equation}
It is then straightforward to obtain
\begin{eqnarray}
\bar u_{m_f} (p') \gamma_\mu u_{m_i} (p) {\cal U}^\mu &=&
C \biggl\{ \left[ (1+\alpha\alpha' {\bf n}' \cdot {\bf n} ) {\cal U}^0
- {\bf a} \cdot \bbox{\cal U} \right] \delta_{m_f\,m_i}
\nonumber \\
&& \quad \mbox{} + i
\left[ \alpha\alpha' ( {\bf n}' \times {\bf n} ) {\cal U}^0 -
( {\bf b} \times \bbox{\cal U} ) \right] \cdot \langle m_f | \bbox{\sigma}
| m_i \rangle \biggr\},
\end{eqnarray}
where we have used ${\bf a} \equiv \alpha' {\bf n}' + \alpha {\bf n}$,
${\bf b} \equiv \alpha' {\bf n}' - \alpha {\bf n}$, and
$ C = \sqrt{(\gamma_p+1)(\gamma_p'+1)}/2$.

With $\Gamma^{\alpha,\mu\nu}$ of Eq. (\ref{Gamalmunu}),
${\cal U}^\alpha$ can be decomposed into
\begin{equation}
{\cal U}^\alpha = 2 ( {\cal V}^\alpha - {\cal W}^\alpha ),
\end{equation}
with
\begin{eqnarray}
{\cal V}^\alpha =  k^\alpha
( \varepsilon_\phi^* \cdot \varepsilon_\gamma ),
\qquad\qquad
{\cal W}^\alpha = ( \varepsilon_\phi^* \cdot k )
\varepsilon_\gamma^\alpha,
\end{eqnarray}
where the identity $\bar u(p') \gamma_\mu u(p) (p-p')^\mu =0$ has been
used in order to simplify the form of ${\cal V}^\alpha$.
Then the Wigner--Eckart theorem enables us to write the VDM amplitude
as the sum of the spin-conserving part and the spin-flip part as in
Eq. (\ref{hel-VDM}).

Using explicit forms of boson polarization vectors in the c.m. frame,
we further have
\begin{equation}
{\cal V}^\mu
= - k^\mu d^{(1)}_{\lambda_\gamma,\lambda_\phi} (\theta)
[ 1 + (\gamma_\phi-1) \delta_{\lambda_\phi\,0} ],
\end{equation}
and
\begin{equation}
{\cal W}^\mu = \left\{ \begin{array}{ll}
\bbox{(} 0,
{\textstyle\frac12} |{\bf k}| \sin\theta,
\pm {\textstyle\frac{i}{2}} |{\bf k}| \sin\theta,
0
\bbox{)} \quad &
\lambda_\phi = \pm 1, \lambda_\gamma = \pm 1,
\\[2mm]
\bbox{(} 0,
- {\textstyle\frac12} |{\bf k}| \sin\theta,
\pm {\textstyle\frac{i}{2}} |{\bf k}| \sin\theta,
0
\bbox{)}
\quad &
\lambda_\phi = \pm 1, \lambda_\gamma = \mp 1,
\\[2mm] \displaystyle
\bbox{(}
0,
\frac{\mp 1}{\sqrt2 M_\phi} |{\bf k}| ( |{\bf q}| - q_0 \cos\theta ),
\frac{-i}{\sqrt2 M_\phi}  |{\bf k}| ( |{\bf q}| - q_0 \cos\theta ),
0
\bbox{)}
\quad &
\lambda_\phi = 0, \lambda_\gamma = \pm 1,
\end{array}
\right.
\nonumber \\
\end{equation}
where $\gamma_\phi \equiv E_\phi^L / M_\phi $, which is close to 1 in the
kinematical range considered in this paper.

\section{Knockout Amplitude for Configuration (III)}

In this case the knockout amplitude is proportional to the matrix element
\begin{eqnarray}
T^{uud\, \rm (III)}_{fi}
\propto i A^* b^{\rm (III)}_1 \sum_{i=1,2,3}
\langle q [s\bar s]^1_{m_{\phi}} \vert
 \langle {p'}[uud]^{1/2}_{m_f} \vert
j^{(i)}_\mu\,\varepsilon^\mu_\gamma
 | \left[ \psi_{J_c}\otimes [s\bar s]^1\right]^{1/2}_{m_i}
 \rangle.
\end{eqnarray}
For simplicity we consider only the  $uud$-quark configuration with the
orbital excitation ($\ell = 1$) of the $[^2{\bf 8,\,{70}}]$ multiplet
with $s_{uud}=\frac12$.
Thus, we do not consider the configuration of $[^4{\bf 8,\,{70}}]$
with $s_{uud}=\frac32$  because the corresponding electromagnetic
transitions are suppressed by the Moorhouse selection rule.
The relevant $uud$-cluster wave function reads
\begin{eqnarray}
\psi_{j m_c} = \langle
{\textstyle\frac{1}{2}}\, m\, 1\, \nu \vert j_c\, m_c \rangle
\frac12 \left[
\psi^{MS}_{\ell,\nu}\left(\phi^{MS}\chi^{MS}-\phi^{MA}\chi^{MA} \right)
+\psi^{MA}_{\ell,\nu}\left(\phi^{MS}\chi^{MA}+\phi^{MA}\chi^{MS} \right)
\right],
\label{APD}
\end{eqnarray}
where $\psi^g_{\ell,\nu}$ is the radial wave function, and $\phi^g$ and
$\chi^g$ are the flavor and spin wave functions respectively.
The superscript $g$ ($=MS,\,MA$) specifies the symmetric property of the
state with respect to the permutation $1 \leftrightarrow  2$.
Using  the Jacobian coordinates of Ref. \cite{Close}, i.e.,
$\xi_2\sim x_1-x_2$ and $\xi_2\sim (x_1+x_2)/2-x_3$, each part of
Eq. (\ref{APD}) reads:
\begin{eqnarray}
&&\psi^{MS}_{1} = \psi_1(\xi_1)\,\psi_0(\xi_2), \qquad
\psi^{MA}_{1} = \psi_0(\xi_1)\,\psi_1(\xi_2),   \nonumber\\
&&\phi^{MS} = \frac{1}{\sqrt{6}}(udu+duu-2uud), \qquad
\phi^{MA} = \frac{1}{\sqrt{2}}(udu-duu),\nonumber\\
&&\chi^{MS}_{\frac12} = \frac{1}{\sqrt{6}}(\uparrow\downarrow\uparrow
+\downarrow\uparrow\uparrow-2\uparrow\uparrow\downarrow),  \qquad
\chi^{MA}_{\frac12} = \frac{1}{\sqrt{2}}(\uparrow\downarrow\uparrow
-\downarrow\uparrow\uparrow).
\end{eqnarray}
The 3-quark proton wave function in this convention has the form
\begin{eqnarray}
\psi_{\frac12, m_p}=
\frac{1}{\sqrt2}\left[
\psi^{S}_{\ell=0}\left(\phi^{MS}\chi^{MS}+\phi^{MA}\chi^{MA} \right)
\right].
\end{eqnarray}

Then it is straightforward to show that the magnetic transition matrix
element between the two $uud$ states is
\begin{eqnarray}
\mbox{M.E.} &=& \langle \psi_{\frac12,m_f} \vert \sum_{j=1,2,3}
\frac{e_j \bbox{\sigma}_j }{2M_u} e^{-ik \cdot x_j}
\vert\psi_{j_c,m_c} \rangle
\cdot ({\bf k}\times \bbox{\varepsilon}_\gamma)
\nonumber\\
&=&
\frac{1}{2M_u}
\langle
\psi_{\frac12,m_f}\vert e_3\bbox{\sigma}_3 e^{-ik \cdot x_3}
\vert \psi_{j_c,m_c} \rangle \cdot({\bf k}\times \bbox{\varepsilon}_\gamma),
\end{eqnarray}
which leads to
\begin{eqnarray}
\mbox{M.E.} &=&
-i\frac{3E_\gamma^L}{2M_u}\frac{\lambda_\gamma\sqrt{3}}{2\sqrt{2}}
\sum
 \langle {\textstyle \frac{1}{2}}\, m\, 1\,\nu \vert j_c\, m_c  \rangle\,
 \langle {\textstyle \frac{1}{2}}\, m\, 1\, \lambda_i \vert
{\textstyle \frac{1}{2}}\, m_f  \rangle\, \langle \psi_{\ell=0}^S \vert
e^{-ik \cdot x_3} \vert\psi_{\ell=1,\nu}^{MS} \rangle
\nonumber\\
&& \mbox{} \times
\langle \phi^{MS}\chi^{MS}+\phi^{MA}\chi^{MA}
\vert e_3\sigma_3^z \vert \phi^{MS}\chi^{MS}-\phi^{MA}\chi^{MA} \rangle.
\end{eqnarray}
Then the spin-flavor part of the matrix element can be evaluated to be
$2e/3$.
The spatial matrix element can be calculated using the standard techniques
of the RHOQM, e.g., as in Ref. \cite{TYO97}.

%%%%%%%%%%%%%%%%%% References %%%%%%%%%%%%%%%%%%%%%%%%%%%%%

%%%%%%%%%%%%%%%%%%%%  Table  %%%%%%%%%%%%%%%%%%%%%%%%%%%%%%

\begin{table}
\begin{center}
\begin{tabular}{|c|c|c|c|c|c|c|}
Configuration & \multicolumn{2}{c|}{I}& \multicolumn{2}{c|}{II}&
\multicolumn{2}{c|}{III}  \\ \hline
$ S_{\bar s s} $    &  0   &  1  &  0  &  1   & 0    &  1  \\ \hline
$s\bar s$-knockout  &  M   & --- & --- & E,M  & ---  & --- \\ \hline
$uud$-knockout      & ---  & E,M & --- & ---  & ---  & E,M \\
\end{tabular}
\end{center}
\caption{Selection rules of knockout processes for each quark configuration
of the nucleon.
The electric and magnetic transitions are represented by E and M,
respectively.}
\label{tab1}
\end{table}

%%%%%%%%%%%%% Fig.1 %%%%%%%%%%%%%%%%%%%%%%%%%%%%%%%%%%%%%%%

%%%%%%%%%%%%%%%%%%%% Fig.2 %%%%%%%%%%%%%%%%%%%%%%%

%%%%%%%%%%%%%%%%%%%%%%% Fig.3 %%%%%%%%%%%%%%%%%%%%%%%%%%%%

%%%%%%%%%%%%%%%%%%%%%%% Fig.4 %%%%%%%%%%%%%%%%%%%%%%%%%%%%

%%%%%%%%%%%%%%%%%%%%% Fig.5 %%%%%%%%%%%%%%%%%%%%%%%%

%%%%%%%%%%%%%%%%%%%% Fig.6 %%%%%%%%%%%%%%%%%%%%%%%%%%%%%

%%%%%%%%%%%%%%%%%%%%%%%% Fig.7 %%%%%%%%%%%%%%%%%%%%%%%%%%

%%%%%%%%%%%%%%%%%%%%%%%%%%%% Fig.8 %%%%%%%%%%%%%%%%%%%%%%%%%%%%%%

%%%%%%%%%%%%%%%%%%%%%%%%% Fig.9 %%%%%%%%%%%%%%%%%%%%%%%%%%%%%%%%%%%%%%%%

%%%%%%%%%%%%%%%%%%%%%%%%%%%% Fig.10 %%%%%%%%%%%%%%%%%%%%%%%%%5

%%%%%%%%%%%%%%%%%%%%%%%%%%%%%%% Fig.11 %%%%%%%%%%%%%%%%%%%%
\begin{figure}
\centering
\epsfig{file=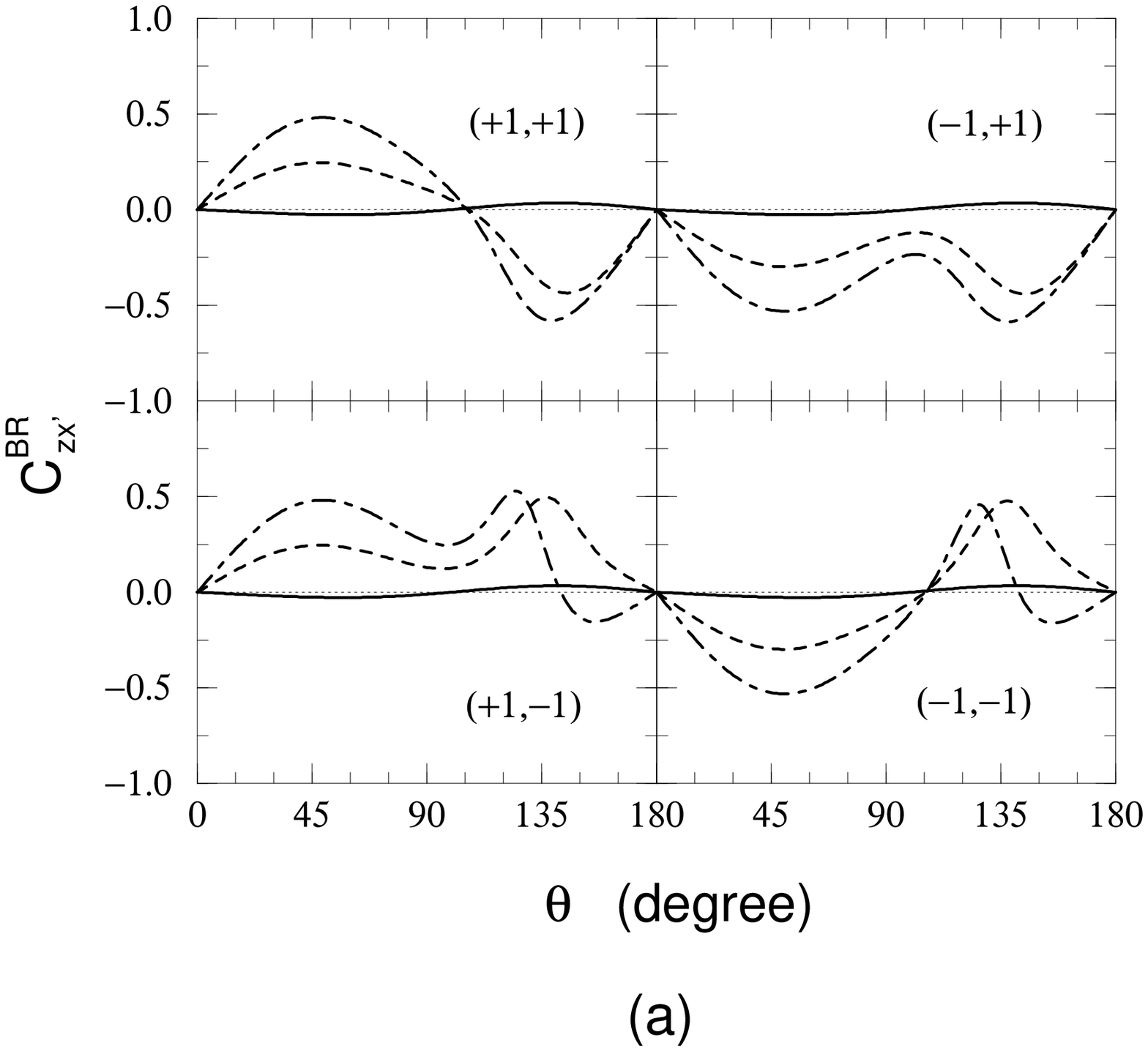, width=7cm} \qquad\qquad
\epsfig{file=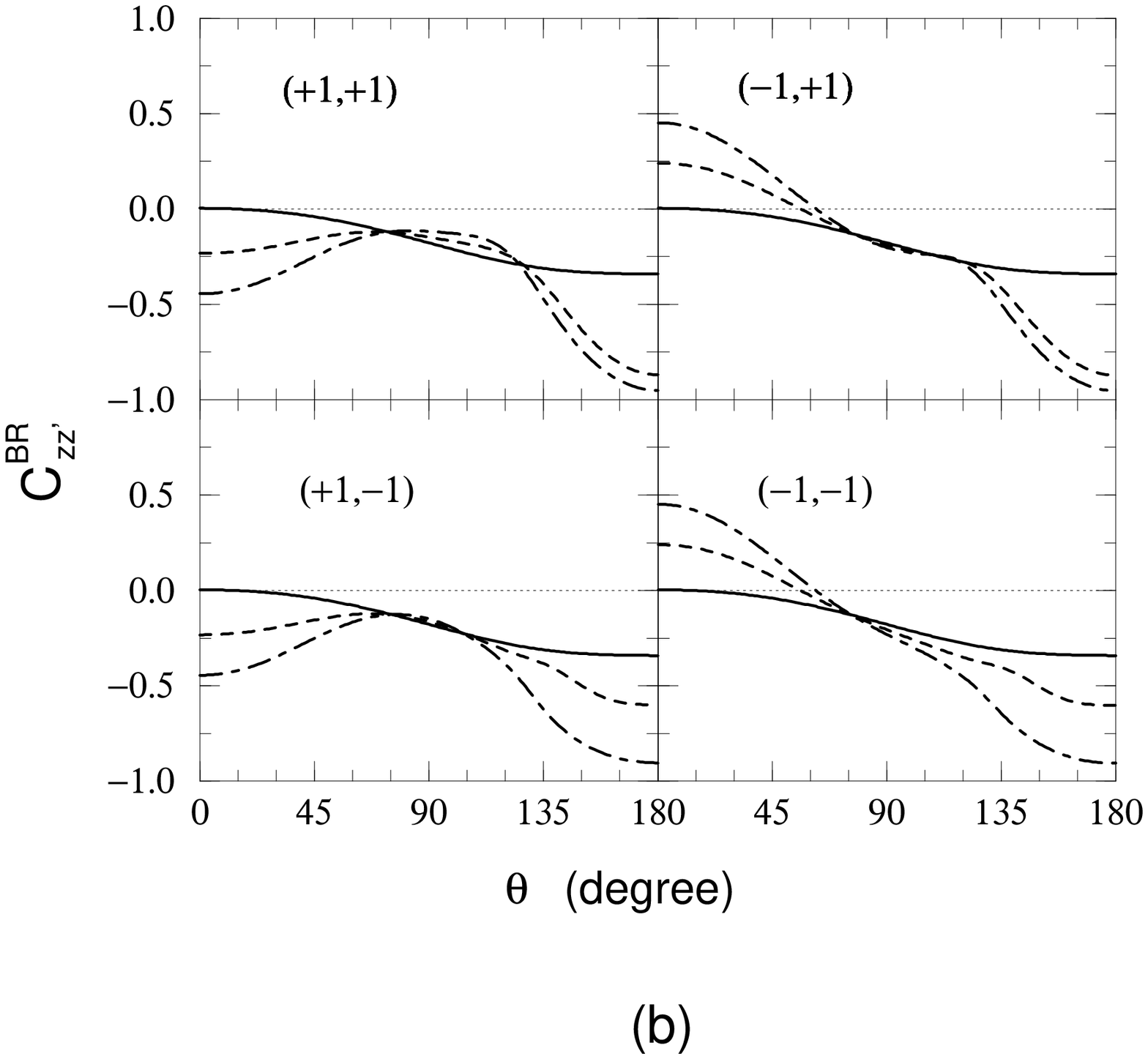, width=7cm}
\caption{Notation same as in Fig. \protect\ref{fig:BT} but for
(a) $C^{\rm BR}_{zx'}$ and (b) $C^{\rm BR}_{zz'}$.}
\label{fig:BR}
\end{figure}

%%%%%%%%%%%%%%%%%%%%%% Fig.12 %%%%%%%%%%%%%%%%%%%%%%%%%%%%%%%%%

\clearpage

%%%%%%%%%%%%%%%%%%%%%%%%%%%%%%% Fig.13 %%%%%%%%%%%%%%%%%%%%%%%%%

%%%%%%%%%%%%%%%%%%%%%%%%%%%% Fig.14 %%%%%%%%%%%%%%%%%%%%%%%%

\clearpage

%%%%%%%%%%%%%%%%%%%%%%%% Fig.15 %%%%%%%%%%%%%%%%%%%%%%%%%%%%%%%

%%%%%%%%%%%%%%%%%%%%%%%% Fig.16 %%%%%%%%%%%%%%%%%%%%%%%%%%%%%%%

%%%%%%%%%%%%%%%%%%%%%%%% Fig.17 %%%%%%%%%%%%%%%%%%%%%%%%%%%%%%%

\end{document}